\documentclass[aip,jcp,reprint,floatfix]{revtex4-2}
\usepackage{color}
\usepackage{graphicx,amsmath}
\usepackage{hyperref}
\hypersetup{colorlinks=true, citecolor=blue, urlcolor=blue, linkcolor=blue}

\newcommand{\beq}{\begin{equation}}
\newcommand{\eeq}{\end{equation}}

\newcommand{\etal} {\textit{et al.}}
\newcommand{\mrm}{\mathrm}
\newcommand{\unit}{\,\mrm}
\newcommand{\rL}{\rho_L}
\newcommand{\muL}{\mu_L}
\newcommand{\PL}{P_L}
\newcommand{\fL}{f_L}
\newcommand{\VL}{V_L}
\newcommand{\rV}{\rho_V}
\newcommand{\muV}{\mu_V}
\newcommand{\PV}{P_V}
\newcommand{\fV}{f_V}
\newcommand{\VV}{V_V}
\newcommand{\SV}{S_V}
\newcommand{\Peq}{P^\infty}
\newcommand{\rav}{\rho_0}
\newcommand{\dav}{\delta_0}
\newcommand{\ds}{\delta_\mathrm{0,sp}}
\newcommand{\de}{\delta_\mathrm{0,eq}}
\newcommand{\xs}{x_\mathrm{sp}}
\newcommand{\xe}{x_\mathrm{eq}}
\newcommand{\dL}{\delta_L}

\newcommand{\thc}{\theta_\mrm{c}}
\newcommand{\SLM}{S_\mathrm{LM}}
\newcommand{\SVM}{S_\mathrm{VM}}
\newcommand{\SLV}{S_\mathrm{LV}}
\newcommand{\rLV}{r_\mathrm{LV}}
\newcommand{\hLV}{h_\mathrm{LV}}
\newcommand{\hLM}{h_\mathrm{LM}}

\newcommand{\Th}{T_h}

\newcommand{\TM}{T_M}
\newcommand{\kB}{k_B}
\newcommand{\kb}{k_B}

\newcommand{\degreeC}{^\circ\mathrm{C}} 

\begin{document}
\title{Effects of compressibility and wetting on the liquid-vapor transition in a confined fluid}

\author{Fr\'ed\'eric Caupin}
\affiliation{Universit\'e de Lyon, Universit\'e Claude Bernard Lyon 1, CNRS, Institut Lumi\`ere Mati\`ere, F-69622, Villeurbanne, France}
\email{frederic.caupin@univ-lyon1.fr}

\date{\today}

\begin{abstract}
When a fluid is constrained to a fixed, finite volume, the conditions for liquid-vapor equilibrium are different from the infinite volume or constant pressure cases. There is even a range of densities for which no bubble can form, and the liquid at a pressure below the bulk saturated vapor pressure remains indefinitely stable. As fluid density in mineral inclusions is often derived from the temperature of bubble disappearance, a correction for the finite volume effect is required. Previous works explained these phenomena, and proposed a numerical procedure to compute the correction for pure water in a container completely wet by the liquid phase. Here we revisit these works, and provide an analytic formulation valid for any fluid and including the case of partial wetting. We introduce the Berthelot-Laplace length $\lambda=2\gamma\kappa/3$, which combines the liquid isothermal compressibility $\kappa$ and its surface tension $\gamma$. The quantitative effects are fully captured by a single, non-dimensional parameter: the ratio of $\lambda$ to the container size.
\end{abstract}

\maketitle

\section{Introduction\label{sec:intro}}

Thermodynamic effects on phase transitions in small systems are well known. The Kelvin equation governs the increase in saturated vapor pressure around a small liquid droplet~\cite{skinner_kelvin_1972}. Free or supported nanocrystals melt at a temperature lower than the bulk, with a larger depression for smaller sizes~\cite{baletto_structural_2005}. When the fluid interacts with the walls of its container, wetting effects come into play and affect phase separation~\cite{gelb_phase_1999}. For the liquid-vapor transition, they are responsible for capillary condensation in small pores and the associated hysteresis~\cite{gelb_phase_1999,huber_soft_2015}. For the liquid-solid transition, melting and freezing hysteresis in pores is often observed, with the pore freezing temperature usually decreasing according to the Gibbs-Thomson law~\cite{gelb_phase_1999,alba-simionesco_effects_2006}. The structure of complex fluids such as liquid crystals is also affected by confinement in nanoporous materials~\cite{huber_soft_2015}.

When in addition the container holding the fluid is closed, new effects are observed. The freezing and melting temperatures of nanoclusters embedded in a matrix differ from those of the bulk or supported material, depending on how the liquid wets the container walls. For instance, germanium nanoclusters, embedded in silica by ion beam implantation, exhibit a strong freezing-melting hysteresis, with phase change temperatures around 930 and 1400 K, that is $\pm 17\%$ around the bulk melting point at 1211.4 K~\cite{xu_large_2006}. This can be explained with a thermodynamic model including wetting effects~\cite{xu_large_2006,caupin_melting_2008}. In the case of germanium, the observations are consistent with a $\pi/2$ contact angle of the liquid-solid interface on the silica substrate~\cite{xu_large_2006,caupin_comment_2007a,xu_xu_2007,caupin_melting_2008}.

In the present work, we are interested in the effect of confinement in a closed container on the liquid-vapor transition. This configuration is known as a Berthelot tube. Indeed, in 1850, Marcelin Berthelot~\cite{berthelot_sur_1850} sealed water in glass tubes. By heating the tubes, thermal expansion of the liquid made the initial vapor bubble disappear. The temperature at which this occurs, called the homogeneization temperature $\Th$, can be translated into the average fluid density in the container. Interestingly, when the tube is cooled, the bubble does not reappear, which shows that the liquid is put under tension (negative pressure): it is stretched to a density less than the saturated liquid density at the same temperature. Later on, this technique has been widely used (see Ref.~\onlinecite{caupin_cavitation_2006} for a review), culminating in record negative pressure down to $-140\unit{MPa}$ when using \textit{microscopic} Berthelot tubes~\cite{roedder_metastable_1967,zheng_limiting_2002,alvarenga_elastic_1993,shmulovich_experimental_2009,el_mekki_azouzi_coherent_2013,pallares_anomalies_2014,pallares_equation_2016,qiu_exploration_2016,holten_compressibility_2017}, i.e. water trapped in micron-sized inclusions in a mineral, usually quartz.
 
An interesting effect is that, in small containers such as the microscopic Berthelot tubes, when a bubble is present, the liquid-vapor equilibrium does not follow the usual thermodynamic path for the bulk material. The bubble curvature induces a pressure difference between liquid and vapor, and the compressibility of the liquid causes the liquid pressure to be lower than the saturated vapor pressure, and even negative~\cite{marti_effect_2012}. This means that a liquid at absolute negative pressure in contact with its vapor can be observed for indefinite periods of time. In some cases, an homogeneous liquid at negative pressure may even be absolutely stable, and thus persists without the threat of nucleation of a vapor bubble (cavitation)\cite{marti_effect_2012}, a phenomenon which has been named \textit{superstability}\cite{wilhelmsen_communication_2014,wilhelmsen_evaluation_2015,vincent_statics_2017}. Moreover, the energy cost associated with the liquid-vapor interface causes a premature bubble instability with respect to the homogeneous stretched liquid. The system homogeneizes at a temperature $\Th$ lower than the value $\Th^\infty$ which would be observed in the absence of surface tension effects~\cite{marti_effect_2012}. This is problematic in the field of geosciences, where $\Th$ of fluid inclusions in minerals is used to determine the fluid density and gain insight on the conditions at which the inclusion was formed~\cite{fall_effect_2009}. For instance, $\Th$ in inclusions formed near the Earth's surface in salt crystals (halite)~\cite{roberts_paleotemperatures_1995,lowenstein_paleotemperatures_1998,guillerm_restoring_2020} or in speleothems made of calcite in caves~\cite{kruger_liquid_2011} serve for reconstructing past temperatures. This motivated Marti~\etal~\cite{marti_effect_2012} to develop their thermodynamic model and propose (for pure water) a procedure to correct the observed $\Th$ and recover $\Th^\infty$, which they applied to speleothems~\cite{spadin_technical_2015}.

We revisit here these phenomena, adding several aspects. Section~\ref{sec:fluid} presents a generic model for the fluid and useful approximations which make our approach applicable to any specific liquid. Section~\ref{sec:wet} retrieves the results of previous works~\cite{marti_effect_2012,wilhelmsen_communication_2014,wilhelmsen_evaluation_2015,vincent_statics_2017}, but thanks to the chosen model, the calculations become fully tractable and give analytic results, and simple, yet accurate approximations. In particular, we introduce in Section~\ref{sec:BL} the Berthelot-Laplace length which controls the fluid behavior, and we show in Section~\ref{sec:Th} how our approach provides a correction procedure equivalent to Ref.~\onlinecite{marti_effect_2012}, but easily applicable to any fluid; as an example, we treat the case of a saturated sodium chloride solution. In Section~\ref{sec:partial}, we add another parameter to the problem, taking into account partial wetting of the liquid on the container walls. In the small bubble limit, the corresponding results can be obtained by a simple rescaling of the Berthelot-Laplace length by a factor calculated from the contact angle. Section~\ref{sec:conclusion} provides a discussion and perspectives.

\section{Generic fluid model\label{sec:fluid}}

A full treatment of the problem requires an explicit equation of state (EoS) for the fluid. For instance, the case of pure water was addressed in Ref.~\onlinecite{marti_effect_2012} using the detailed EoS from the International Association for the Properties of Water and Steam (IAPWS)~\cite{theinternationalassociationforthepropertiesofwaterandsteam_revised_2018,wagner_iapws_2002}, valid for both the liquid and vapor phase in a broad temperature and pressure range. However, far from the liquid-vapor critical point, the vapor is not dense and can be approximated by a perfect gas. Furthermore, the narrow range of relevant liquid densities allows the use of a generic liquid EoS which makes the results directly applicable to any liquid whose surface tension $\gamma$ and compressibility $\kappa$ are known. 

\subsection{Lowest order equation of state for the liquid\label{sec:eos}}

In this section, we consider the liquid at a fixed temperature $T$: all parameters are thus defined at this temperature. Quantities corresponding to liquid-vapor equilibrium of an infinite system with a flat interface are indicated with the superscript $\infty$.

\begin{figure}
\centering
\includegraphics[width=0.95\columnwidth]{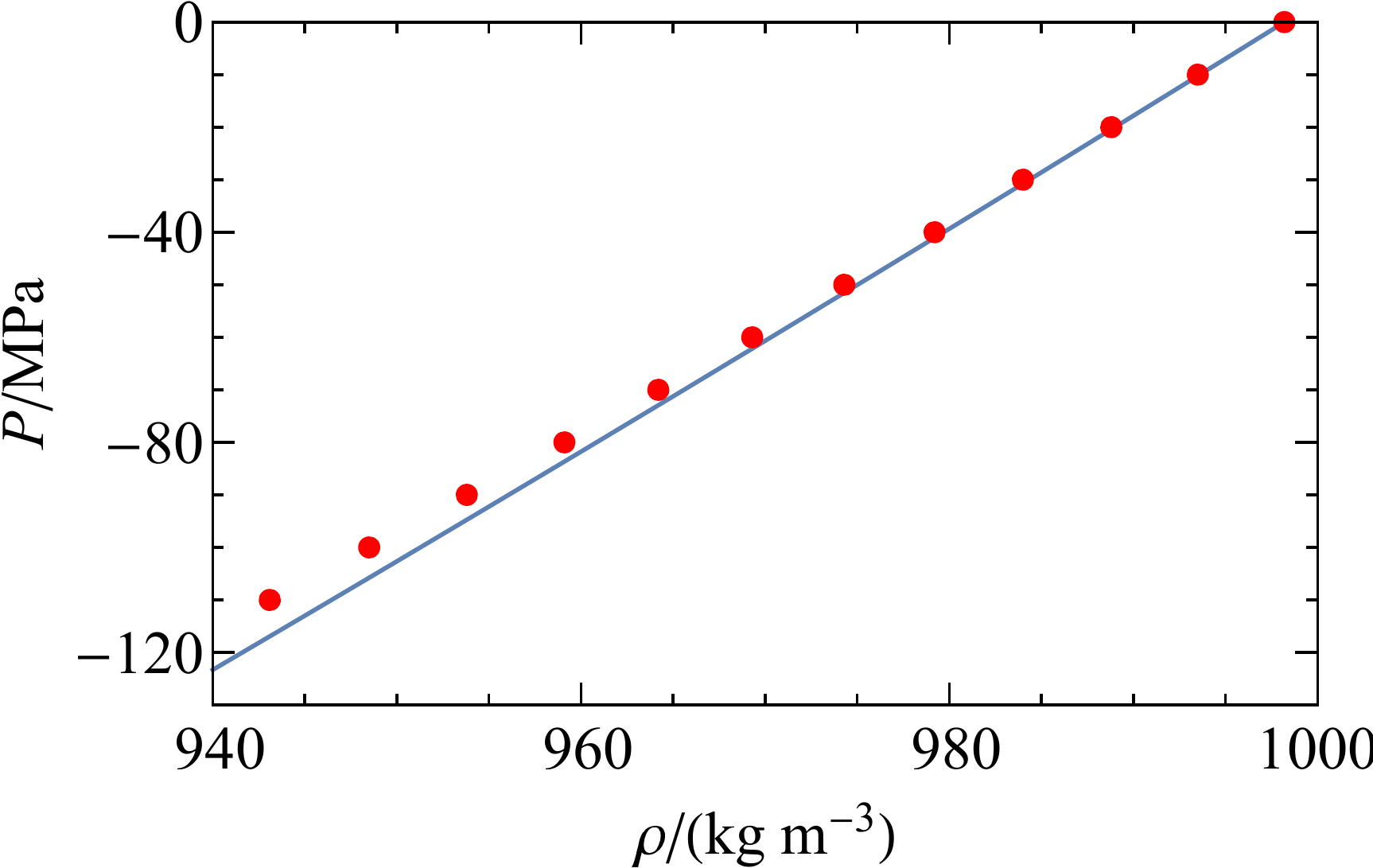}
\caption{(color online): Pressure as a function of density for pure water at $20\,\degreeC$. The red discs show the experimental EoS at negative pressure\cite{pallares_equation_2016}, and the blue curve Eq.~\ref{eq:P}.
\label{fig:EoS}
}
\end{figure}

Our starting point is a linear expansion in density $\rL$ of the chemical potential $\mu_L$:
\beq
\mu_L = \muL^\infty + \frac{1}{{\rL^\infty}^2 \kappa} (\rL - \rL^\infty ) \, ,
\label{eq:mu}
\eeq
where $\kappa$ is the liquid isothermal compressibility at saturated vapor pressure. The Gibbs-Duhem relation $(\partial \mu/\partial P)_T = 1/\rho$ gives the following dependence on $\rL$ for the pressure $P$:
\beq
\PL = \Peq + \frac{1}{2 \kappa} \left[ \left( \frac{\rL}{\rL^\infty}\right)^2 -1 \right] \, ,
\label{eq:P}
\eeq
where $\Peq = \PL^\infty = \PV^\infty$. Figure~\ref{fig:EoS} shows that Eq.~\ref{eq:P} matches well experimental data at $20\,\degreeC$ down to $-50\,\mathrm{MPa}$ at least. Note that the expansion leading to Eq.~\ref{eq:P} is different from those based on constant $\partial \rho/\partial P$ or constant isothermal compressibility used in Refs.~\onlinecite{wilhelmsen_communication_2014,wilhelmsen_evaluation_2015} and~\cite{vincent_statics_2017}, respectively. The two last generate a logarithmic dependence in the chemical potential (Eq.~\ref{eq:mu}), or in the pressure (Eq.~\ref{eq:P}), respectively. With the present choice instead, the liquid isothermal compressibility $\kappa_T$ varies with density as $\kappa_T (\rL ) = (\rL^\infty / \rL )^2 \kappa$. Logarithms are avoided and the resolution is simplified, allowing analytic expressions to be obtained as detailed below.

The Helmholtz free energy per unit volume is:
\beq
f_i = \rho_i \mu_i - P_i \, ,
\label{eq:f}
\eeq
where the subscript $i = L$ or $V$ for the liquid and vapor, respectively.

\subsection{Approximations far from the critical point\label{sec:approx}}

As we are dealing with a fluid at temperatures far below the liquid-vapor critical temperature, we will make simplifying assumptions. We will assume $\rV  \ll \rL$, and treat the vapor as a perfect gas. In the canonical ensemble, the total volume $V$ is fixed: $V=\VL + \VV$, where $\VL$ and $\VV$ are the volumes of the liquid and vapor phases, respectively. The total number of particles $N$ is also fixed:
\beq
N = \rho_0 V = \rL \VL + \rV \VV \, ,
\label{eq:N}
\eeq
where $\rho_0 = N/V$ is the fixed average density of the system. Let us introduce the following reduced quantities:
\beq
\dav = \frac{\rho_0}{\rL^\infty}, \quad \dL = \frac{\rL}{\rL^\infty}, \quad \mathrm{and} \quad x=\frac{\VV}{V} \, .
\label{eq:red}
\eeq
Eqs.~\ref{eq:mu} and \ref{eq:P} then rewrite:
\beq
\muL (\rL)= \muL^\infty + \frac{1}{\rL^\infty \kappa} (\dL - 1 ) \, ,
\label{eq:muL}
\eeq
\beq
\PL (\rL)= \Peq + \frac{{\dL}^2-1}{2 \kappa} \, .
\label{eq:PL}
\eeq

We will only consider dense systems with high $\rav$, so that $\VV \ll \VL$. With $\rV  \ll \rL$, Eq.~\ref{eq:N} becomes:
\beq
\dL = \frac{\dav}{1-x} \, .
\label{eq:dL}
\eeq

One of the governing equations of the problem is the equality of chemical potentials, $\muL (\rL) = \muV(\rV)$ (see Section~\ref{sec:Df}). Using Eq.~\ref{eq:mu} and the perfect gas equation $\PV = \rV \kB T$, this gives:
\beq
\muL^\infty + \frac{1}{{\rL^\infty}^2 \kappa} (\rL - \rL^\infty ) = \muV^\infty + \kB T \ln \frac{\PV}{\PV^\infty} \, ,
\label{eq:muLmuV}
\eeq
which leads to:
\beq
\PV = \Peq \exp \left( \frac{\dL -1}{\kB T \rL^\infty \kappa} \right)
\label{eq:PV}
\eeq

In the cases we study, $|\dL - 1| \ll \kB T \rL^\infty \kappa$; for instance, the latter term is $\simeq 0.6$ for water at ambient conditions, whereas $1-\dL$ will be a few $10^{-3}$ at most. This allows a further simplification:
\beq
\PV \simeq \Peq + \frac{\Peq}{\kB T \rL^\infty \kappa} \left( \dL -1 \right)
\label{eq:PVapprox}
\eeq
\beq
\PL (\rL) - \PV = \frac{{\dL}^2-1}{2 \kappa} - \frac{\Peq}{\kB T \rL^\infty \kappa} \left( \dL -1 \right)
\label{eq:DP}
\eeq

In the right hand side, the ratio of the second term to the first is:
\beq
\frac{2\Peq}{\kB T \rL^\infty (\dL +1)} = \frac{2 \rV^\infty}{\rL^\infty} \frac{1}{\dL +1} \leq \frac{\rV^\infty}{\rL^\infty} \ll 1 \, ,
\label{eq:ratio}
\eeq
so that within the same degree of approximation as the rest, we may write:
\beq
\PL (\rL) - \PV \simeq \frac{{\dL}^2-1}{2 \kappa} \, .
\label{eq:DPapprox}
\eeq

\subsection{Capillarity approximation\label{sec:cap}}

To describe the liquid-vapor interface, we will use the capillarity approximation: the interface is considered as infinitely thin, separating two regions of space each having a constant density, and its energy per unit area is given by the bulk equilibrium surface tension $\gamma$. This is valid when the bubble radius is large compared to the physical thickness of the liquid-vapor interface, around $1\,\mathrm{nm}$ for water near ambient conditions~\cite{caupin_liquid-vapor_2005}. In that case, it is also safe to neglect the variation of surface tension with curvature, as the Tolman length is typically a fraction of the molecular size~\cite{bruot_curvature_2016}. When the bubble radius becomes comparable to the interfacial thickness, more elaborate treatments are required, such as density functional theory~\cite{caupin_liquid-vapor_2005}.

\section{Fluid confined in a container completely wet by the liquid\label{sec:wet}}

In this section we consider a fluid whose liquid phase completely wets the container walls. This means that, when a vapor bubble is present, there is no contact between the vapor and the wall. The container with volume $V$ may be of any shape, but for convenience we introduce its typical size $R$ as the radius of the sphere with the same volume: $V =(4/3) \pi R^3$. The system being closed and isothermal, we consider the canonical ensemble: $N$ particles in a fixed volume $V$ at constant temperature $T$. 

\subsection{Free energy change in the canonical ensemble\label{sec:Df}}

The thermodynamic potential of the system is the Helmholtz free energy $F$. Its reference value $F_0$ corresponds to the homogeneous liquid at $\rL = N \rav = N/V$: $F_0 = \fL ( \rav ) V$. When the system contains a bubble of volume $\VV$ and surface area $\SV$, the liquid volume and density change to $\VL = V - \VV $ and to $\rL$, respectively. The free energy change writes:
\beq
\Delta F = \fL ( \rL ) \VL + \fV ( \rV ) \VV + \gamma \SV - \fL (\rav ) V \, .
\label{eq:DF1}
\eeq

For a fixed bubble (constant $\VV$ and $\SV$), minimizing $\Delta F$ with respect to $\rL$ under the constraints $\VV + \VL = V$ and $\rL \VL + \rV \VV = \rav V$ yields the condition of equal chemical potentials:
\beq
\mu_L (\rL ) = \mu_V (\rV ) \, .
\label{eq:eqmu}
\eeq 

Using Eqs.~\ref{eq:f},~\ref{eq:N}, and~\ref{eq:eqmu}, we rearrange Eq.~\ref{eq:DF1} into:
\begin{widetext}
\beq
\Delta F = \rav V [\muL (\rL) - \muL (\rav )] + [\PL (\rav) - \PL (\rL )] V + [\PL (\rL) - \PV ] \VV + \gamma \SV \, .
\label{eq:DF2}
\eeq
\end{widetext}

Note that this relation, obtained by minimizing $\Delta F$ with respect to $\rL$ at fixed bubble shape, is independent of the specific EoS of the fluid. For a fixed vapor volume $\VV$, the bubble shape which minimizes $\Delta F$ is the sphere (lowest $\SV$). This shape is allowed here as we assume the liquid to fully wet the container walls.

We now make use of the EoS (Section~\ref{sec:eos}) and of the approximations valid far from the liquid-vapor critical point (Section~\ref{sec:approx}). With Eqs.~\ref{eq:red},~\ref{eq:muL},~\ref{eq:PL},~\ref{eq:dL}, and~\ref{eq:DPapprox}, after some rearrangements, we obtain:
\beq
\Delta F = \frac{\VV}{2 \kappa} \left( \frac{{\dav}^2}{1-x} -1 \right) + \gamma \SV \, .
\label{eq:DF3}
\eeq

Let $r$ be the bubble radius. Introducing the non-dimensional free energy change $\phi = 2 \kappa \Delta F/V$, we get:
\beq
\phi = \left( \frac{{\dav}^2}{1-x} -1 \right) x + \frac{6 \gamma \kappa}{R} x^{2/3} \, .
\label{eq:phi1}
\eeq

\subsection{The Berthelot-Laplace length\label{sec:BL}}

We define the Berthelot-Laplace length $\lambda$ as follows:
\beq
\lambda = \frac{2}{3} \gamma \kappa \, .
\label{eq:BL}
\eeq

The liquid compressibility $\kappa$ is the key to understand how a Berthelot tube works. To justify the name ``Berthelot-Laplace'' and give $\lambda$ a physical meaning, let us consider $N$ particles in a spherical liquid droplet in equilibrium with its vapor. Without surface tension, it would have a radius $R_d$ such that $(4/3) \pi {R_d}^3 \rL^\infty = N$. However, surface tension induces a pressure jump across the liquid-vapor interface, the Laplace pressure $\Delta P = 2 \gamma/R_d$. Because the liquid is compressible, this pressure increase results in a density increase $\delta \rL = \kappa \rL^\infty \Delta P = 2 \gamma \kappa \rL^\infty / R_d$. The radius thus decreases to $R_d + \delta R_d$ such that:
\beq
\frac{4}{3} \pi (R_d + \delta R_d)^3 (\rL^\infty + \delta \rL ) = N = \frac{4}{3} \pi {R_d}^3 \rL^\infty \, ,
\eeq
which gives to first order:
\beq
\frac{\delta R_d}{R_d} = - \frac{2 \gamma \kappa}{3 R_d} = - \frac{\lambda}{R_d} \, .
\label{eq:shrink}
\eeq

The relative decrease in radius is thus given by the ratio of the Berthelot-Laplace length to the droplet radius. The name Berthelot-Laplace signals this physical interpretation, demonstrating the effect of the Laplace pressure on a compressible liquid.

\begin{figure}
\centering
\includegraphics[width=0.95\columnwidth]{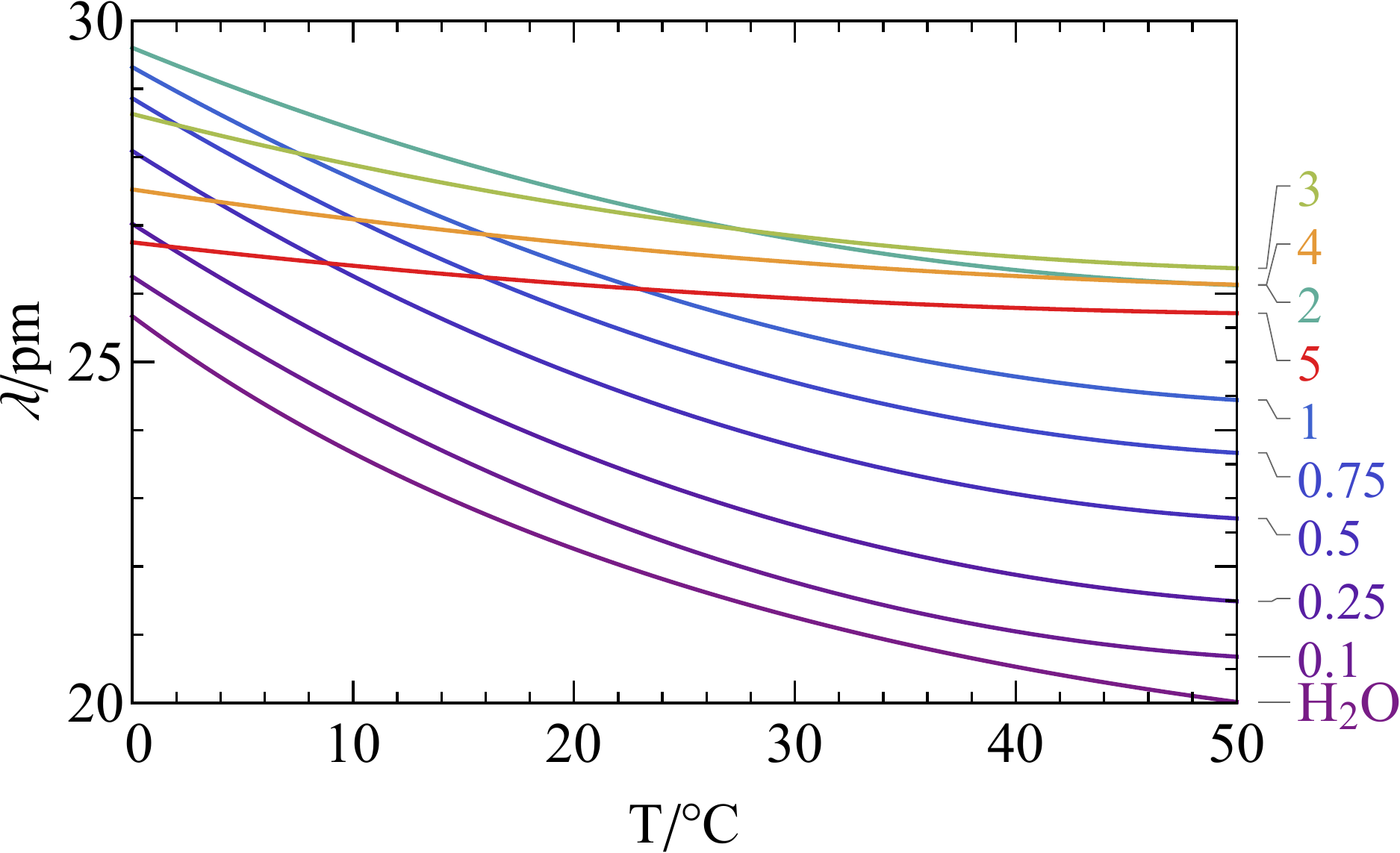}
\caption{(color online): Berthelot-Laplace length as a function of temperature for a series of NaCl solutions. The curves are labeled by the molality in $\mathrm{mol\,kg^{-1}}$.
\label{fig:BL}
}
\end{figure}

Figure~\ref{fig:BL} shows $\lambda$ from 0 to $50\degreeC$ for water and aqueous NaCl solutions. For water, we used the relevant IAPWS releases for isothermal compressibility\cite{theinternationalassociationforthepropertiesofwaterandsteam_revised_2018,wagner_iapws_2002} and surface tension~\cite{theinternationalassociationforthepropertiesofwaterandsteam_revised_2014}. We find the following cubic fit:
\begin{eqnarray}
\lambda(T)/\mathrm{pm} = 25.6407 - 0.228344 \,T&\nonumber\\
+ 0.00337068 \,T^2 - 0.0000212502 &\,T^3 
\label{eq:lambdaH2O}
\end{eqnarray}
with $T$ in $\degreeC$ to represent the data within $0.11\%$. For NaCl solutions, we use $\kappa$ calculated from the Rogers-Pitzer EoS~\cite{rogers_volumetric_1982} with an estimated confidence limit of $0.5\%$, and surface tension calculated with the correlation from Ref.~\onlinecite{dutcher_surface_2010} which fits experimental with an average absolute and maximum percentage error of $0.72\%$~\cite{dutcher_surface_2010} and $1.71\%$~\cite{nayar_surface_2014}, respectively. The expected uncertainty on $\lambda$ is therefore less than $1\%$. When increasing the salt concentration, $\lambda$ first increases, but around a molality of $3\unit{mol\,kg^{-1}}$ the temperature dependence becomes milder and $\lambda$ starts decreasing. The decrease of $\kappa$ when salt is added eventually overcomes the increase in $\gamma$. The curves become flatter because the isothermal compressibility minimum, around $42\degreeC$ in pure water, moves to lower temperatures. In Section~\ref{sec:Th}, we will consider the case of a saturated NaCl solution. Unfortunately, supporting data for $\kappa$ is available only up to $5\unit{mol\,kg^{-1}}$, whereas the saturation molality is above $5\unit{mol\,kg^{-1}}$ (Ref.~\onlinecite{farelo_solidliquid_1993}). An extrapolation gives $\lambda\simeq 26\unit{pm}$, and based on the trend of the curves, we will assume this constant value at all temperatures. We see that, in all cases, $\lambda$ is extremely small, smaller than the molecular dimensions. This shows that the relative radius decrease given by Eq.~\ref{eq:shrink} will become noticeable only for tiny droplets.

\subsection{Stationary points and the Laplace equation\label{sec:Laplace}}

Coming back to the thermodynamics of the confined fluid introduced in Section~\ref{sec:Df}, Eq.~\ref{eq:phi1} rewrites:
\beq
\phi = \left( \frac{{\dav}^2}{1-x} -1 \right) x + 9 \epsilon x^{2/3} \, ,
\label{eq:phi2}
\eeq
where $\epsilon = \lambda/R$. This expression is simpler and more generic than that obtained with an accurate, but complex and specific, EoS, as the IAPWS-95 EoS~\cite{marti_effect_2012}. As mentioned in Section~\ref{sec:eos}, it avoids logarithmic terms, enabling analytic calculations as described in the following. Furthermore, the small values of $\lambda$ will allow us to treat $\epsilon$ as a small quantity.

\begin{figure}
\centering
\includegraphics[width=0.95\columnwidth]{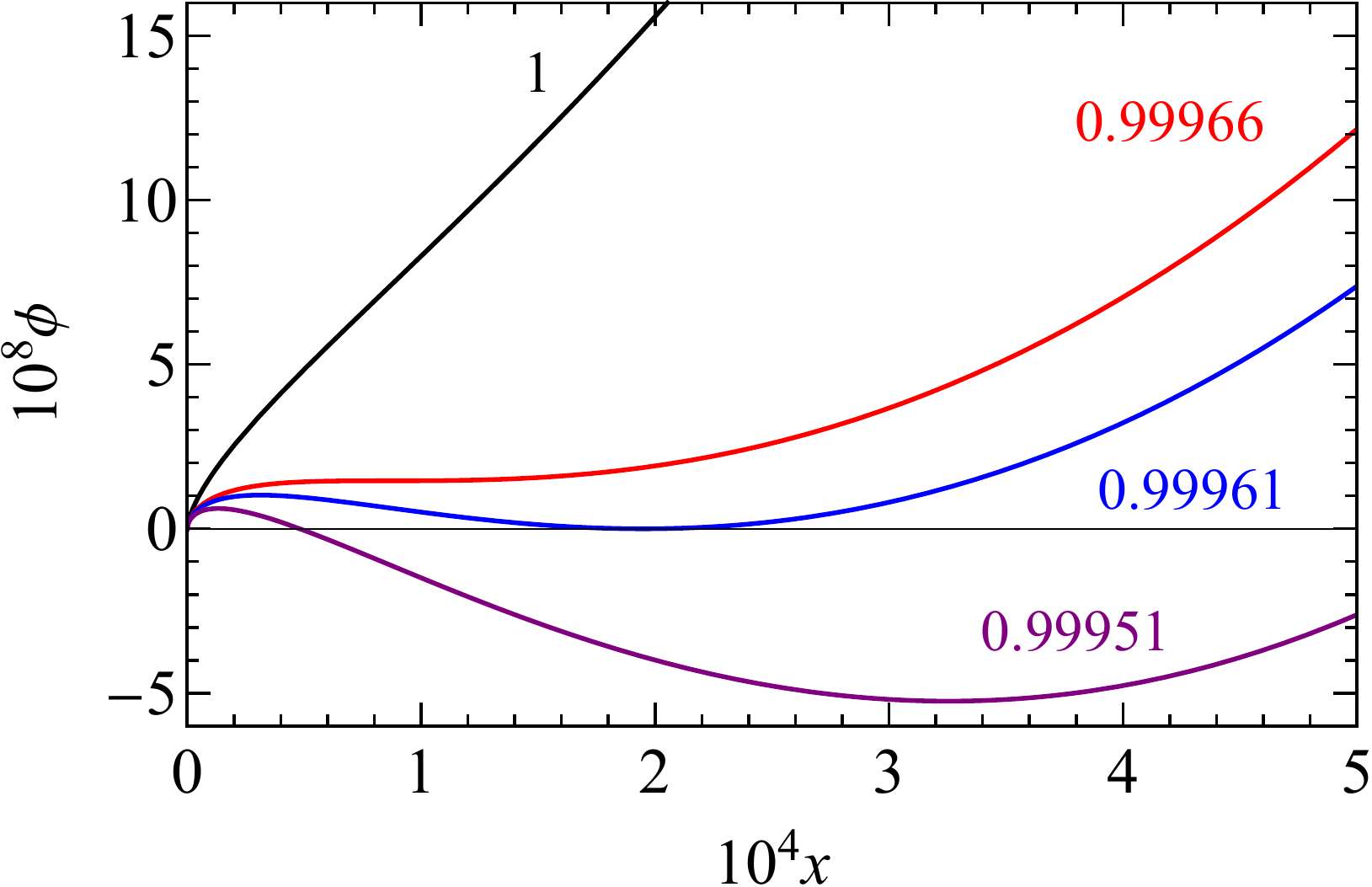}
\caption{(color online): Reduced free energy as a function of bubble volume for pure water at $12\unit{\degreeC}$ and $V=10^3\,\mathrm{\mu m^3}$ ($\epsilon = 3.8\,10^{-6}$). The curves are labeled by $\dav$.
\label{fig:phi}
}
\end{figure}
To see the possible states of the system, it is instructive to plot $\phi (x)$ for various values of $\dav$, as shown in Fig.~\ref{fig:phi}. We consider several degrees of stretching, that is when the average density $\rav$ is less than the saturated liquid density $\rL^\infty$, i.e. $\dav \leq 1$. This analysis has already been performed\cite{marti_effect_2012,wilhelmsen_communication_2014,wilhelmsen_evaluation_2015,vincent_statics_2017}. We repeat it here using Eq.~\ref{eq:phi2}.

At $\dav = 1$, $\phi$ is a monotonically increasing function of $x$: there is no reason for a bubble to form, and if one was created, it would collapse immediately. At low enough $\dav$ on the contrary, a negative minimum develops at a finite $x$, indicating the presence of a stable bubble. The homogeneous liquid state is separated from this minimum by an energy barrier which decreases with decreasing $\dav$.

The stationary points for which $(\partial \phi/\partial x)=0$ fulfill the condition:
\beq
\delta_0 = (1-x) \sqrt{ 1 - \frac{6 \epsilon}{x^{1/3}}} \, .
\label{eq:stat}
\eeq

From Eqs.~\ref{eq:dL}, \ref{eq:DPapprox}, and \ref{eq:stat}, and noticing that $x=(r/R)^3$, one finds for the pressure difference between the vapor and the liquid phases:
\beq
\PV - \PL(\rL) = \frac{3 \epsilon}{2 \kappa x^{1/3}} = \frac{2\gamma}{r} \, ,
\eeq
thus recovering the venerable Laplace equation. This corresponds to mechanical equilibrium of the liquid-vapor interface; this mechanical equilibrium is stable if $\phi$ is minimum, or unstable otherwise. To discuss the overall, thermodynamic stability or metastability of the system, one needs to identify local and global minima of $\phi$. In between the two extreme cases $\rav = 1$ and low $\rav$, the system will pass through ranges of parameters with a qualitatively distinct behavior. The limiting cases separating these ranges are studied in Sections~\ref{sec:spino} and \ref{sec:eq}.

\subsection{Bubble instability\label{sec:spino}}

\begin{figure}
\centering
\includegraphics[width=0.95\columnwidth]{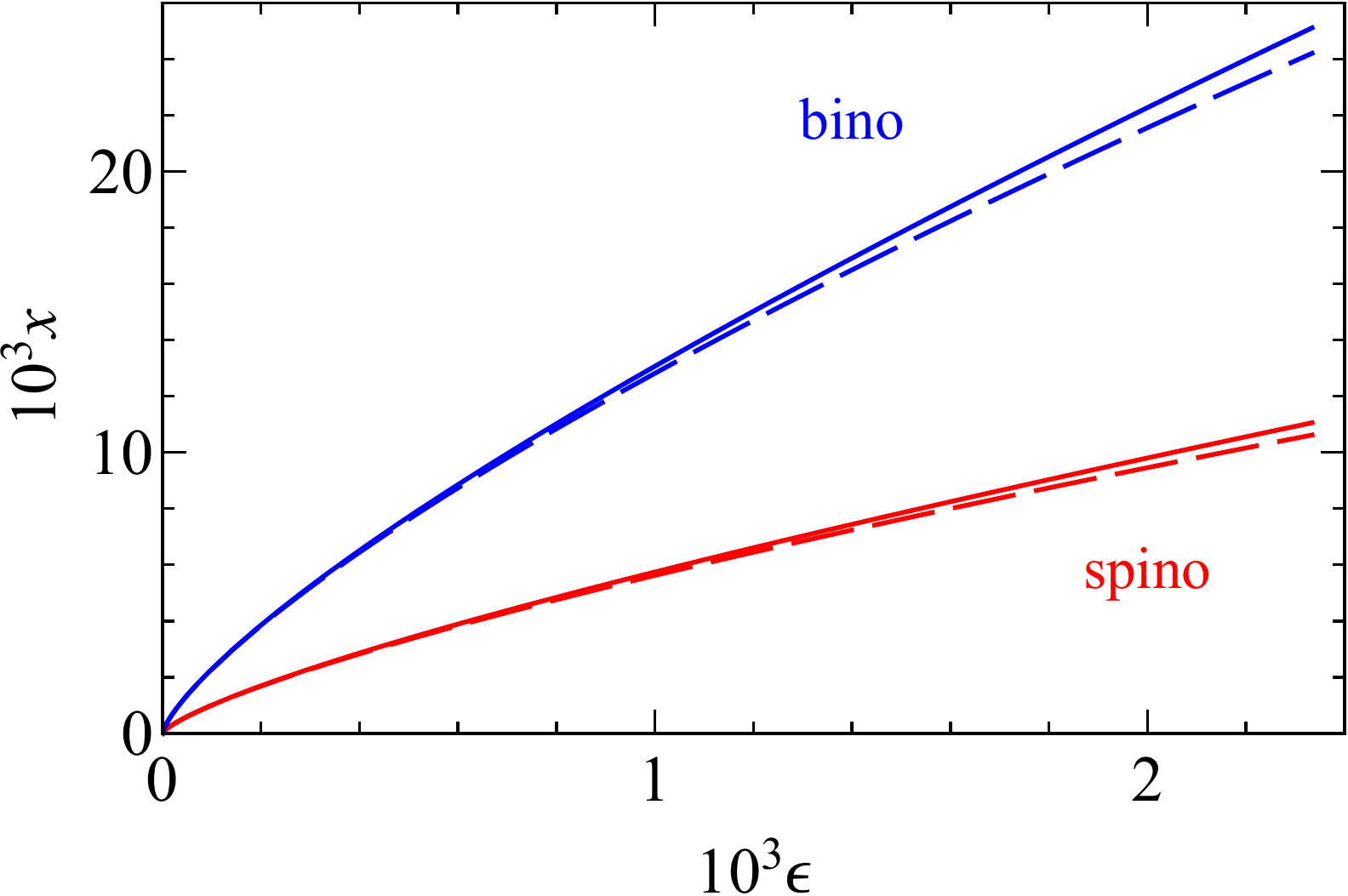}
\caption{(color online): Bubble volume at the bubble spinodal (red) and binodal (blue). The solid and dashed curves show the exact and approximate (Eqs.~\ref{eq:xsapprox} and \ref{eq:xeapprox}) results, respectively.
\label{fig:xse}
}
\end{figure}

\begin{figure}
\centering
\includegraphics[width=0.95\columnwidth]{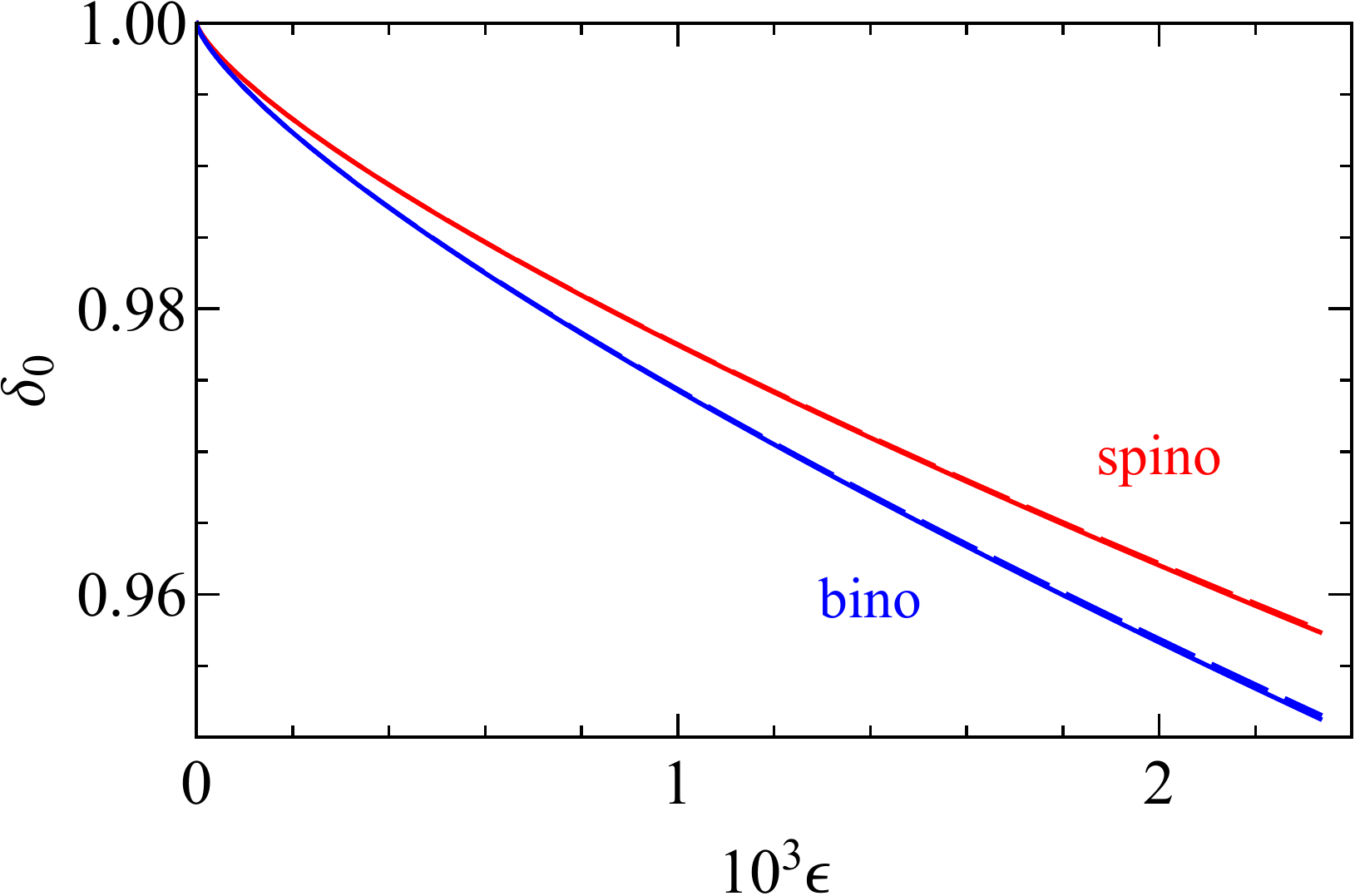}
\caption{(color online): Average density $\dav$ at the bubble spinodal (red) and binodal (blue). The solid and dashed curves show the exact and approximate (Eqs.~\ref{eq:dsapprox} and \ref{eq:deapprox}) results, respectively.
\label{fig:dse}
}
\end{figure}

One first particular value of $\dav$ is that for which $\phi$ develops an inflection point with horizontal tangent. Above this value, no bubble, even metastable, can form. This lead Marti~{\etal} to call this point a \textit{bubble spinodal}. The spinodal value of $\dav$, $\ds$, and that of $x$, $\xs$, are thus obtained from the conditions that $\partial \phi/\partial x$ and $\partial^2 \phi/\partial x^2$ simultaneously vanish. The latter condition gives the following relation:
\beq
\ds = \frac{(1-x)^{3/2} \epsilon^{1/2}}{x^{2/3}}
\label{eq:ds}
\eeq
which, when substituted into Eq.~\ref{eq:stat}, gives a quartic equation for $\xs^{1/3}$:
\beq
{\xs}^{4/3} - 5 \epsilon \xs - \epsilon = 0 \, .
\label{eq:xsquart}
\eeq
This equation admits analytic solutions which are too lengthy to be displayed here. For relevant values of the parameters, two solutions are complex, and two are real with opposite signs. We select the positive solution for $\xs$, and deduce $\ds$ from Eq.~\ref{eq:ds}. The values as a function of $\epsilon$ are displayed in Figs.~\ref{fig:xse} and \ref{fig:dse}.

As $\epsilon \ll 1$, we expand the solutions to lowest order. The results are particularly compact:
\begin{eqnarray}
\label{eq:xsapprox}\xs & = &\epsilon^{3/4} + \mathcal{O}(\epsilon^{3/2}) \, ,\\
\label{eq:dsapprox}\ds & = & 1- 4\epsilon^{3/4} + \mathcal{O}(\epsilon^{3/2}) \, .
\end{eqnarray}
The error made with these approximate formulas decreases with increasing $R$. For water at $12\unit{\degreeC}$ and $R=10\unit{nm}$, $\epsilon = 0.0023$, and the errors on $\xs$ and $1-\ds$ are already less than 4 and 0.4 \%, respectively.

We can also calculate $\PV - \PL(\rL)$, the pressure difference between the vapor and the liquid phases at the bubble spinodal, and compare it to $\Peq - \PL(\rav )$, the tension reached by the liquid in the homogeneous state (without bubble):
\begin{eqnarray}
\frac{\PV - \PL(\rL)}{\Peq - \PL(\rav )} = \frac{1 - \dL^2}{1 - \dav^2} \simeq \frac{1 - \dL}{1 - \dav}& \nonumber \\
= \frac{1-\xs-\ds}{(1-\xs)(1-\ds)}\simeq 1 -& \frac{\xs}{1-\ds} \simeq \frac{3}{4} \, .
\end{eqnarray}
This is indeed what is observed in several examples of pressure paths followed by fluid inclusions in Figs.~4b and 5b of Ref.~\onlinecite{marti_effect_2012}.

Finally, we comment about the criterion previously derived for \textit{superstability}\cite{wilhelmsen_communication_2014,wilhelmsen_evaluation_2015,vincent_statics_2017}. With a different approximation for the EoS, the minimum volume of the container below which no bubble can form is given by Eq.~(8) of Ref.~\onlinecite{wilhelmsen_communication_2014}, which becomes with our notations:
\beq
V_\mathrm{min} = V_\mathrm{V} \left( \frac{3r}{2\gamma\kappa} +1 \right) = V_\mathrm{V} \left( \frac{r}{\lambda} +1 \right) \, .
\label{eq:WBKR}
\eeq
At the bubble spinodal, $\xs=V_\mathrm{V}/V_\mathrm{min}$. In the limit $\lambda \ll r$, Eq.~\ref{eq:WBKR} rewrites:
\beq
\xs\simeq\frac{\lambda}{r}=\frac{\lambda}{R}\frac{R}{r}=\frac{\epsilon}{{\xs}^{1/3}}\, ,
\eeq
which yields the same result as Eq.~\ref{eq:xsapprox}, $\xs \simeq \epsilon^{3/4}$.

\subsection{Liquid-vapor equilibrium\label{sec:eq}}

When $\dav$ becomes lower than $\ds$, a second minimum appears at finite $x$ in $\phi (x)$ with $\phi >0$, indicating the possibility of a metastable bubble. At low enough $\dav$, $\phi$ at this minimum becomes negative, indicating a stable bubble; the homogeneous liquid state becomes in turn metastable. At a value $\de$, the bubble minimum has $\phi=0$ as the homogeneous liquid state: the two states can exist as stable states. This lead Marti~{\etal} to call this point a \textit{bubble binodal}, but emphasizing that only one state can be observed at a time in one container. The \textit{binodal} or \textit{equilibrium} values of $\dav$, $\de$, and that of $x$ $\xe$, are thus obtained from the conditions that $\phi$ and $\partial \phi/\partial x$ simultaneously vanish. Combining Eqs.~\ref{eq:phi2} and \ref{eq:stat} leads to a quartic equation for $\xe^{1/3}$:
\beq
{\xe}^{4/3} - 6 \epsilon \xe - 3 \epsilon = 0 \, .
\label{eq:xequart}
\eeq
This equation admits analytic solutions which are too lengthy to be displayed here. For relevant values of the parameters, two solutions are complex, and two are real with opposite signs. We select the positive solution for $\xe$, and deduce $\de$ from Eq.~\ref{eq:stat}.

As $\epsilon \ll 1$, we expand the solutions to lowest order near 0. The results are again particularly compact:
\begin{eqnarray}
\label{eq:xeapprox}\xe & = &(3 \epsilon)^{3/4} + \mathcal{O}(\epsilon^{3/2}) \, ,\\
\label{eq:deapprox}\de & = & 1- 2 (3 \epsilon)^{3/4} + \mathcal{O}(\epsilon^{3/2}) \, .
\end{eqnarray}
The error made with these approximate formulas decreases with increasing $R$. For water at $12\unit{\degreeC}$ and $R=10\unit{nm}$, $\epsilon = 0.0023$, and the errors on $\xs$ and $1-\ds$ are already less than 3.4 and 0.6 \%, respectively.

We can also calculate $\PV - \PL(\rL)$, the pressure difference between the vapor and the liquid phases at the bubble binodal, and compare it to $\Peq - \PL(\rav )$, the tension reached by the liquid in the homogeneous state (without bubble):
\begin{eqnarray}
\frac{\PV - \PL(\rL)}{\Peq - \PL(\rav )} = \frac{1 - \dL^2}{1 - \dav^2} \simeq \frac{1 - \dL}{1 - \dav} & \nonumber \\
= \frac{1-\xe-\de}{(1-\xe)(1-\de)}\simeq 1 -& \frac{\xe}{1-\de} \simeq \frac{1}{2} \, .
\end{eqnarray}
This is indeed what is observed in several examples of pressure paths followed by fluid inclusions in Figs.~4b and 5b of Ref.~\onlinecite{marti_effect_2012}.

Finally, we comment about the physical meaning of the ``coexistence'' of two states with the same free energy in the system. Although as emphasized by Marti~{\etal}, only one state can be observed at a time in one container, there is a possibility for the system transiting back and forth between the two states, as already suggested for the liquid-solid transition in nanoclusters~\cite{reiss_capillarity_1988}. This may happen inasmuch as the energy barrier separating the two minima is low enough to be overcome by thermal fluctuations. Taking a $50\kB T$ limit and water at $12\unit{\degreeC}$ as in Fig.~\ref{fig:phi} as an example, this requires an inclusion smaller than $0.018\unit{\mu m^3}$, i.e. a radius of $160\unit{nm}$. This is small, but amenable to experiments.

\subsection{Effect on bubble disappearance temperature\label{sec:Th}}

As mentioned in the introduction, a negative consequence of the confinement-compressibility effect is that the actual homogeneization temperature $\Th$ is lower than the value $\Th^\infty$ which would be observed in the absence of surface tension~\cite{marti_effect_2012}. Indeed, when the temperature increases at fixed density $\rav$, the saturated liquid density $\rL^\infty (T)$ decreases, so that there is a temperature $\Th < \Th^\infty$ at which $\dav (T)=\rav/\rL^\infty (T)$ reaches $\ds$ for this temperature. This raises concerns for obtaining the density of fluid inclusions or using them for paleotemperature reconstruction. 

Ref.~\onlinecite{fall_effect_2009} considered this question. Unfortunately, it was erroneously assumed that the pressure difference between the vapor and the liquid phases just before the bubble disappears, $\PV - \PL(\rL)$, was equal to the tension reached by the liquid in the homogeneous state (without bubble), $\Peq - \PL(\rav )$ (Eq.~7 in Ref.~\onlinecite{fall_effect_2009}). Sections~\ref{sec:spino} and \ref{sec:eq} show that $[\PV - \PL(\rL)]/[\Peq - \PL(\rav )]$ is in fact between 1/2 and 3/4. Marti~\etal~\cite{marti_effect_2012} revisited this problem more rigorously and developed a procedure to correct the observed $\Th$ and recover $\Th^\infty$ in the case of pure water.

\begin{figure}
\centering
\includegraphics[width=0.95\columnwidth]{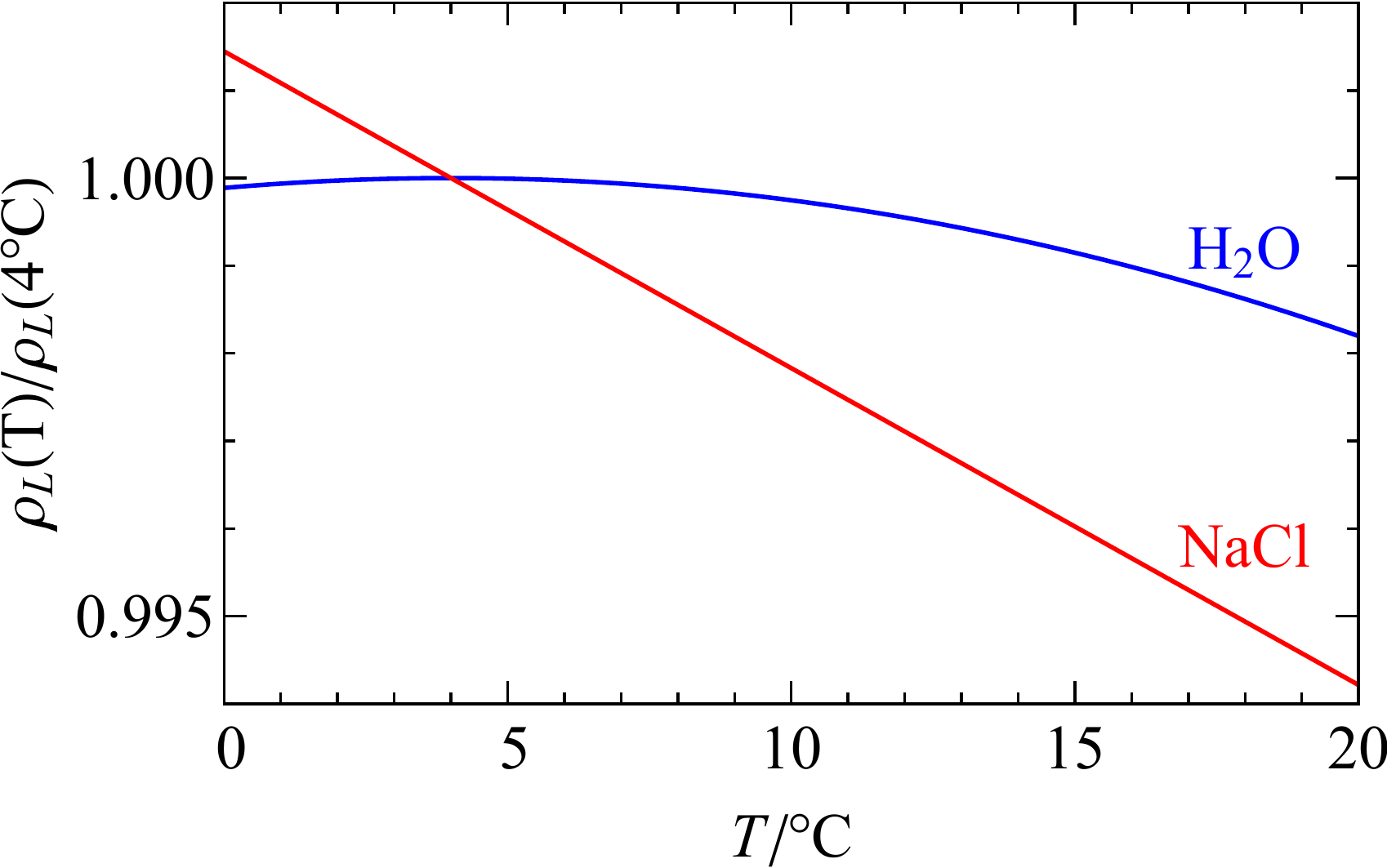}
\caption{(color online): Ratio $\rL^\infty(T)/\rL^\infty(\TM)$ as a function of temperature for pure water (blue) and a saturated NaCl solution (red).
\label{fig:rhosat}
}
\end{figure}

Let us first show how our approach can recover their results. To this aim, we need a model for $\rL^\infty (T)$. We sample data every $0.2\unit{\degreeC}$ from the formula of the IAPWS Revised Supplementary Release on Saturation Properties of Ordinary Water Substance~\cite{theinternationalassociationforthepropertiesofwaterandsteam_revised_1992} which indicates an uncertainty of $0.01\unit{kg\,m^{-3}}$. In view of the modest temperature interval relevant to paleotemperature, we use a simple parabolic expansion around the temperature of maximum density for pure water $\TM=4.003\unit{\degreeC}$:
\beq
\rL^\infty (T) = \rho_L^M \left[ 1 - \alpha \left(\frac{T}{\TM} -1 \right)^2 \right] \, ,
\label{eq:rhoLsat}
\eeq
where $\rL^M=999.922\unit{kg\,m^{-3}}$ is the maximum density of saturated liquid water, and $\alpha=0.541284$ fits the sampled saturation densities from $\TM$ to $20\unit{\degreeC}$ with standard and maximum absolute deviations of $0.014$ and $0.039\unit{kg\,m^{-3}}$, respectively. The ratio $\rL/\rL^M$ is displayed in Fig.~\ref{fig:rhosat}.

Following Marti~\etal~\cite{marti_effect_2012}, we consider only a truly isochoric system (fixed $V$). If necessary, a correction for the thermal expansion of the matrix can be easily incorporated in the model~\cite{spadin_technical_2015}. We first assume $\Th^\infty$ to be known, and calculate the corresponding actual $\Th$. It is comprised between two extreme temperatures: $T_{h,\mathrm{sp}}$ at which $\dav(T_{h,\mathrm{sp}}) = \ds$ (bubble spinodal), and $T_{h,\mathrm{eq}}$ at which $\dav(T_{h,\mathrm{eq}}) = \de$ (bubble binodal). The temperature dependence of $\lambda$ (Fig.~\ref{fig:BL}) requires a numerical resolution of the relevant equations:
\begin{eqnarray}
\label{eq:rhoThs}\rL^\infty(\Th^\infty) &=& \rL^\infty(T_{h,\mathrm{sp}}) \,\ds \left[\frac{\lambda(T_{h,\mathrm{sp}})}{R} \right] \, ,\\
\label{eq:rhoThe}\rL^\infty(\Th^\infty) &=& \rL^\infty(T_{h,\mathrm{eq}}) \,\de \left[\frac{\lambda(T_{h,\mathrm{eq}})}{R} \right] \,.
\end{eqnarray}

\begin{figure}
\centering
\includegraphics[width=0.95\columnwidth]{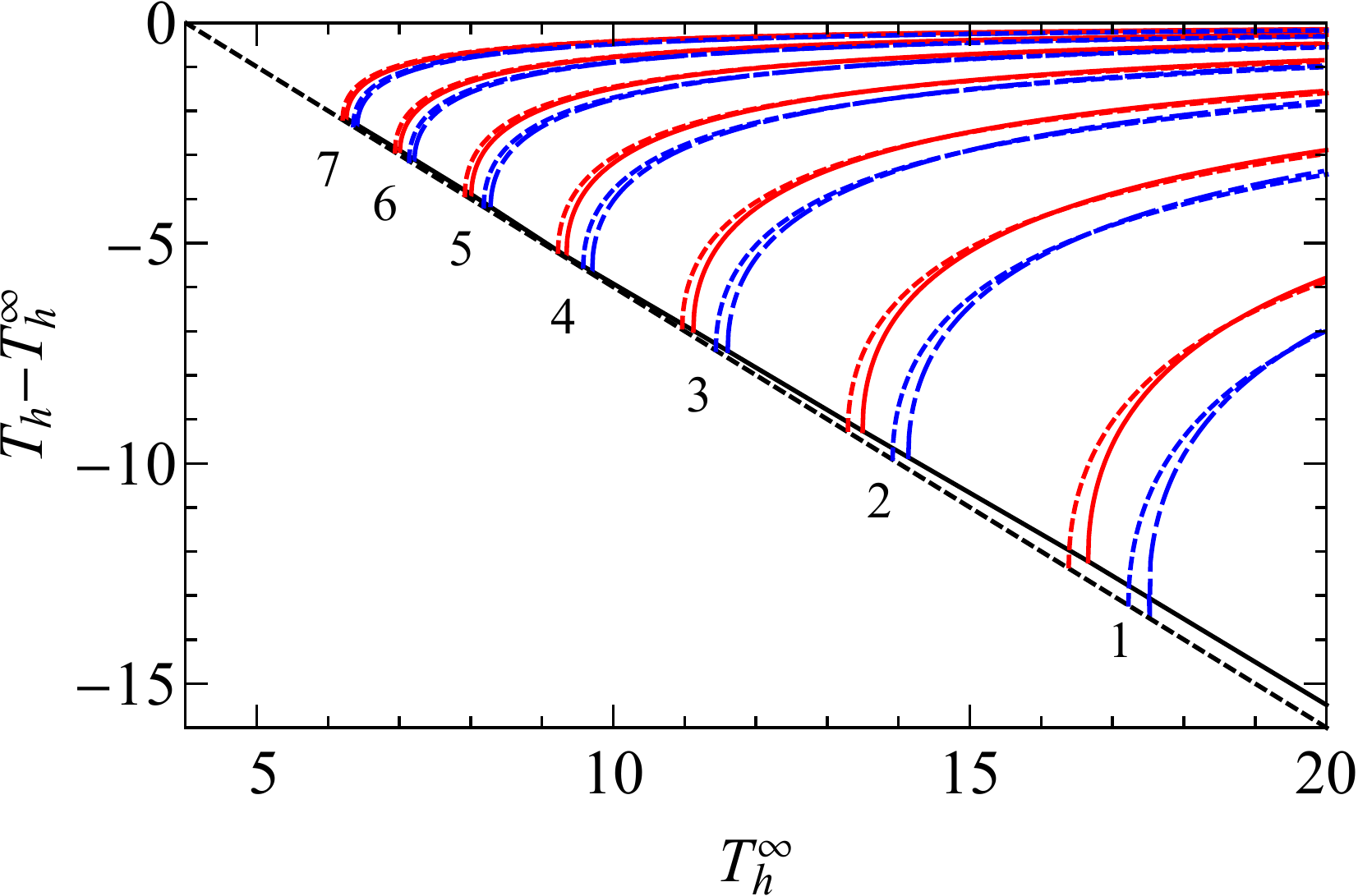}
\caption{(color online): $\Th - \Th^\infty$ as a function of $\Th^\infty$. The labels give $\log[V/(10^3\unit{\mu m^3})]$. The exact solutions for the bubble spinodal and binodal are shown with solid red and dot-dashed blue curves, respectively. The solutions for a constant $\lambda=23.3\unit{pm}$ are shown with dashed curves. Below the diagonal lines no bubble can exist.
\label{fig:DT}
}
\end{figure}

The numerical solutions are used to plot $\Th^\infty - \Th$ as a function of $\Th^\infty$ for a series of inclusion volumes in Fig.~\ref{fig:DT}, which is identical to Fig.~8b of Ref.~\onlinecite{marti_effect_2012}.

Still, it is interesting to approximate $\lambda$ by a constant value, because it restores analytic results; from Eqs.~\ref{eq:ds}, \ref{eq:stat}, and \ref{eq:rhoLsat}, we find:
\begin{eqnarray}
\label{eq:Ths}T_{h,\mathrm{sp}} &=& \TM \left[ 1+ \sqrt{\left(\frac{\Th^\infty}{\TM} -1 \right)^2 - \frac{4 }{\alpha}\left(\frac{\lambda}{R}\right)^{3/4}} \right]\\
\label{eq:The}T_{h,\mathrm{eq}} &=& \TM \left[ 1+ \sqrt{\left(\frac{\Th^\infty}{\TM} -1 \right)^2 - \frac{2 }{\alpha}\left(\frac{3\lambda}{R}\right)^{3/4}} \right]\, .
\end{eqnarray}

Figure~\ref{fig:DT} shows that the simple approximations obtained when replacing $\lambda(T)$ by its value at $12\degreeC$ agree well with the numerical solutions. As noted in Ref.~\onlinecite{marti_effect_2012}, there is an area in which no bubble can exist. With the approximate solution, this occurs exactly when $\Th=\TM$, whereas for the exact solution, it occurs at $\Th \gtrsim \TM$ because $\lambda(T)$ is a decreasing function of temperature.

Marti~{\etal} then proceed with an interesting argument to provide a procedure to correct $\Th$ experimentally. The idea is to measure both the vapor bubble radius $r$ at $\TM$ and the actual $\Th$, which, combined with the model, yields $\Th^\infty$. More precisely, it yields an interval for $\Th^\infty$, depending on whether the bubble disappears at the spinodal ($\Th = T_\mathrm{sp}$) or the binodal ($\Th = T_\mathrm{eq}$) temperature.

The equilibrium bubble radius at $\TM$ provides $\dav(\TM)$ through Eq.~\ref{eq:stat}. On the other hand, to find $\dav(T_{h,\mathrm{sp}})$ and $\dav(T_{h,\mathrm{eq}})$, we have simple formulas for $\ds$ (Eq.~\ref{eq:ds}) and $\de$ (Eq.~\ref{eq:stat}). 
Noticing that $x=(r/R)^3$ and $\dav(\TM)/\dav(T) = \rL^\infty(T)/\rL^\infty(\TM)$, we arrive at two quartic equations on $(1/R)^{3/4}$:
\begin{widetext}
\begin{eqnarray}
\label{eq:Rs}\left[ 1- \left( \frac{r(\TM)}{R} \right)^3 \right] \sqrt{1 - 6 \frac{\lambda(\TM)}{r(\TM)}} &=& \left[1 - \alpha \left(\frac{\Th}{\TM} -1  \right)^2\right] \left[1-4\left(\frac{\lambda(\Th)}{R}\right)^{3/4}\right]\quad\mathrm{for}\quad\Th=T_\mathrm{sp}\, ,\\
\label{eq:Re}\left[ 1- \left( \frac{r(\TM)}{R} \right)^3 \right] \sqrt{1 - 6 \frac{\lambda(\TM)}{r(\TM)}} &=& \left[1 - \alpha \left(\frac{\Th}{\TM} -1  \right)^2\right] \left[1-2\left(\frac{3\lambda(\Th)}{R}\right)^{3/4}\right]\quad\mathrm{for}\quad\Th=T_\mathrm{eq}\, .
\end{eqnarray}
\end{widetext}

These equations admit analytic solutions which are too lengthy to be displayed here. For relevant values of the parameters, two solutions are complex, and two are real with opposite signs. We select the positive solutions for $R$ ($R_\mathrm{sp}$ from Eq.~\ref{eq:Rs} and $R_\mathrm{eq}$ from Eq.~\ref{eq:Re}), and deduce the limits on $\Th^\infty$ from Eqs.~\ref{eq:Ths} and ~\ref{eq:The}, respectively:
\begin{eqnarray}
T_\mathrm{h,min}^\infty =& \nonumber\\
\TM &\left[ 1+ \sqrt{\left(\frac{\Th}{\TM} -1 \right)^2 + \frac{4}{\alpha} \left(\frac{\lambda(\Th)}{R_\mathrm{sp}}\right)^{3/4}} \;\right]\, ,\label{eq:Thinfs}\\
T_\mathrm{h,max}^\infty =& \nonumber\\
\TM &\left[ 1+ \sqrt{\left(\frac{\Th}{\TM} -1 \right)^2 + \frac{2}{\alpha} \left(\frac{3\lambda(\Th)}{R_\mathrm{eq}}\right)^{3/4}}\;\right]\, .\label{eq:Thinfe}
\end{eqnarray}

\begin{figure}
\centering
\includegraphics[width=0.95\columnwidth]{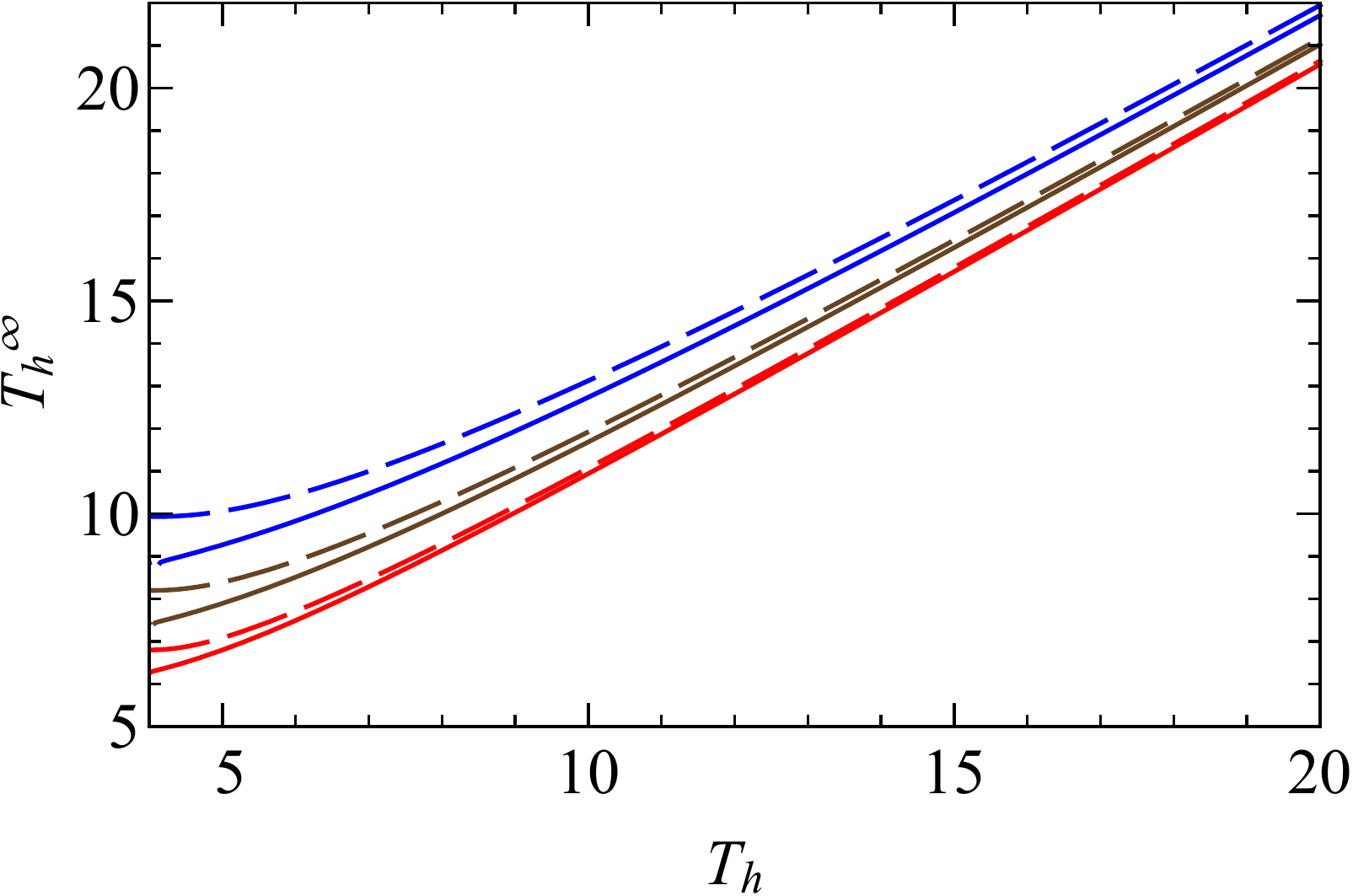}
\caption{(color online): $\Th^\infty$ as a function of observed $\Th$ for three bubble radii at $\TM$. From top to bottom: $r=0.6$, $1.2$, and $2.7\unit{\mu m}$. The solutions for the bubble spinodal and binodal are shown with solid and dashed curves, respectively. \label{fig:Thinf}
}
\end{figure}

We use these expressions to plot $\Th^\infty $ as a function of $\Th$ for three bubble radii $r$ at $\TM$, and obtain Fig.~\ref{fig:Thinf} which is identical to the inset in Fig.~9a of Ref.~\onlinecite{marti_effect_2012}. We note that Marti~{\etal} indicated that this calculation of $\Th^\infty$ could not be done analytically, whereas our approach allows obtaining analytic expressions.

\begin{figure}
\centering
\includegraphics[width=0.95\columnwidth]{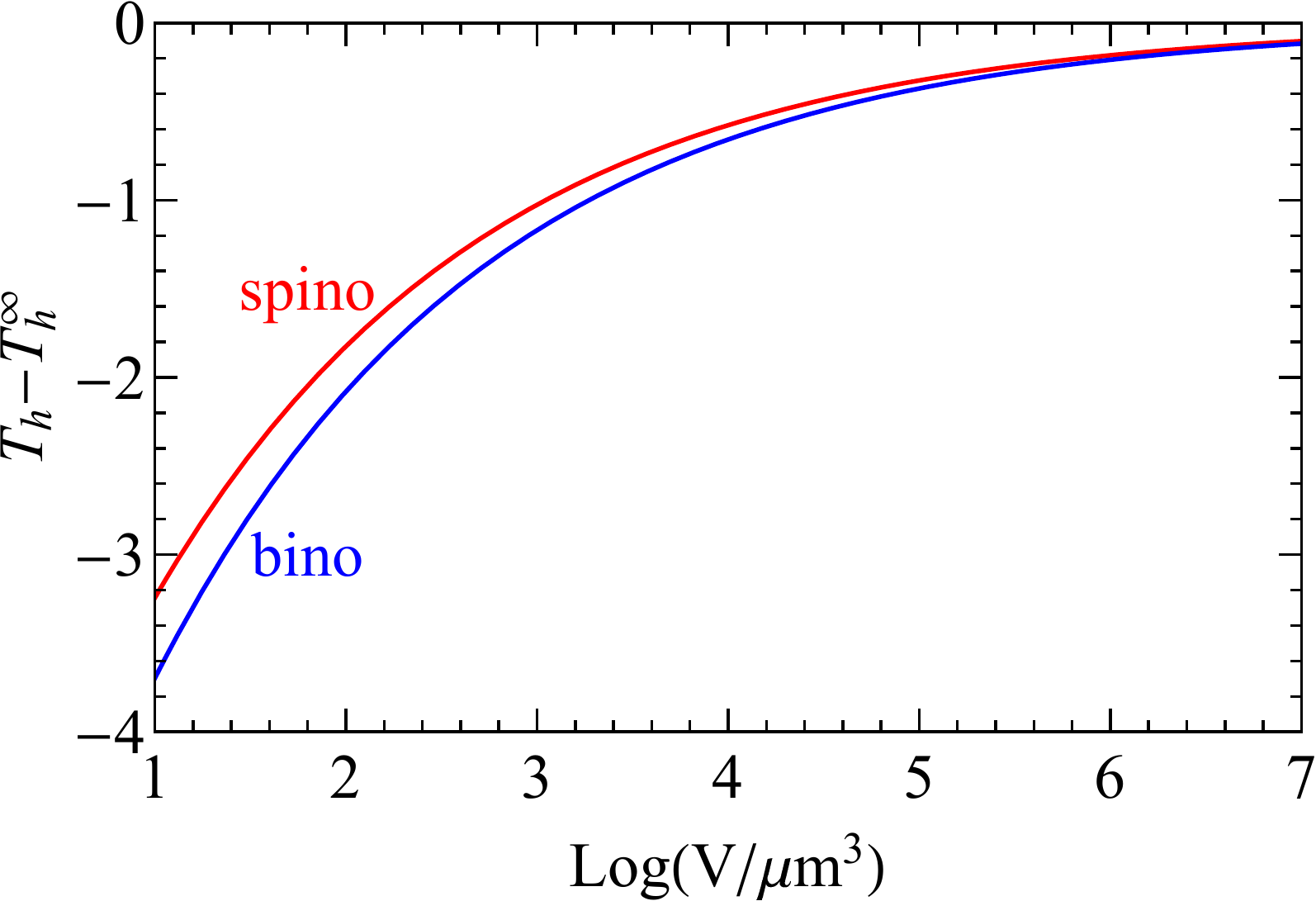}
\caption{(color online): $\Th - \Th^\infty$ as a function of volume $V$ for a saturated NaCl solution at the bubble spinodal (red) and binodal (blue).
\label{fig:DTNaCl}
}
\end{figure}

Our general approach allows us to repeat the same analysis with saturated NaCl solution. We note that for saturated NaCl solution, $\lambda(T)$ is nearly constant (see Section~\ref{sec:BL}), which makes the calculations simpler. We also need to replace Eq.~\ref{eq:rhoLsat} with the temperature dependence of the density of a saturated NaCl solution at saturated vapor pressure $P^\infty(T)$. To this aim, we use the EoS of Ref.~\onlinecite{alghafri_densities_2012} and compute the density at $0.1\unit{MPa}$ (neglecting the difference with $P^\infty(T)$) at the saturation molality given by Farelo~{\etal}~\cite{farelo_solidliquid_1993}; the latter is given above $20\unit{\degreeC}$ but extrapolates smoothly to $0\unit{\degreeC}$. This gives
\beq
\rho_{L,\mathrm{sat. NaCl}}^\infty (T) = a - b T\, ,
\label{eq:rhoLsatNaCl}
\eeq
with $a=1209.05\mathrm{kg\,m^{-3}}$ and $b=0.436565\unit{kg\,m^{-3}\,\degreeC^{-1}}$, $T$ being in $\degreeC$. This linear behavior, displayed in Fig.~\ref{fig:rhosat}, varies much faster than the parabolic function for pure water. This changes the results drastically. Solving Eqs.~\ref{eq:Ths} and \ref{eq:The} with the NaCl parameters gives:
\begin{eqnarray}
T_{h,\mathrm{sp}} - \Th^\infty &=& \frac{4\epsilon^{3/4}}{1-4\epsilon^{3/4}} \left(\Th^\infty - \frac{a}{b} \right)\nonumber\\
&\simeq& - 4\epsilon^{3/4} \frac{a}{b} = -4\left(\frac{\lambda}{R}\right)^{3/4}\frac{a}{b} \, ,\label{eq:ThsNaCl}\\
T_{h,\mathrm{eq}} - \Th^\infty &=& \frac{2(3\epsilon)^{3/4}}{1-2(3\epsilon)^{3/4}} \left(\Th^\infty - \frac{a}{b} \right)\nonumber\\
&\simeq& - 2(3\epsilon)^{3/4} \frac{a}{b} = -2\left(\frac{3\lambda}{R}\right)^{3/4}\frac{a}{b} \, ,\label{eq:TheNaCl}
\end{eqnarray}
where the relations $\epsilon \ll 1$ and $\Th^\infty \ll a/b$ have allowed further simplification. The decrease in homogeneization temperature for the brine is therefore much simpler than for pure water, becoming independent of $\Th^\infty$ itself, and being only a weak function of the inclusion volume $V$. Figure~\ref{fig:DTNaCl} displays this function.

The residual volume dependence still requires to correct the observed $\Th$ using a bubble radius measurement to obtain $\Th^\infty$. For simplicity we assume that the bubble radius is still measured at $\TM$ of pure water, although for the brine there is no $\rL$ maximum any more. Equations~\ref{eq:Rs} and \ref{eq:Re} are replaced by two new quartic equations on $(1/R)^{3/4}$:
\begin{widetext}
\begin{eqnarray}
\label{eq:RsNaCl}\left[ 1- \left( \frac{r(\TM)}{R} \right)^3 \right] \sqrt{1 - 6 \frac{\lambda}{r(\TM)}} &=& \frac{a-b\Th}{a-b\TM} \left[1-4\left(\frac{\lambda}{R}\right)^{3/4}\right]\quad\mathrm{for}\quad\Th=T_\mathrm{sp}\, ,\\
\label{eq:ReNaCl}\left[ 1- \left( \frac{r(\TM)}{R} \right)^3 \right] \sqrt{1 - 6 \frac{\lambda}{r(\TM)}} &=& \frac{a-b\Th}{a-b\TM} \left[1-2\left(\frac{3\lambda}{R}\right)^{3/4}\right]\quad\mathrm{for}\quad\Th=T_\mathrm{eq}\, .
\end{eqnarray}
\end{widetext}

\begin{figure}
\centering
\includegraphics[width=0.95\columnwidth]{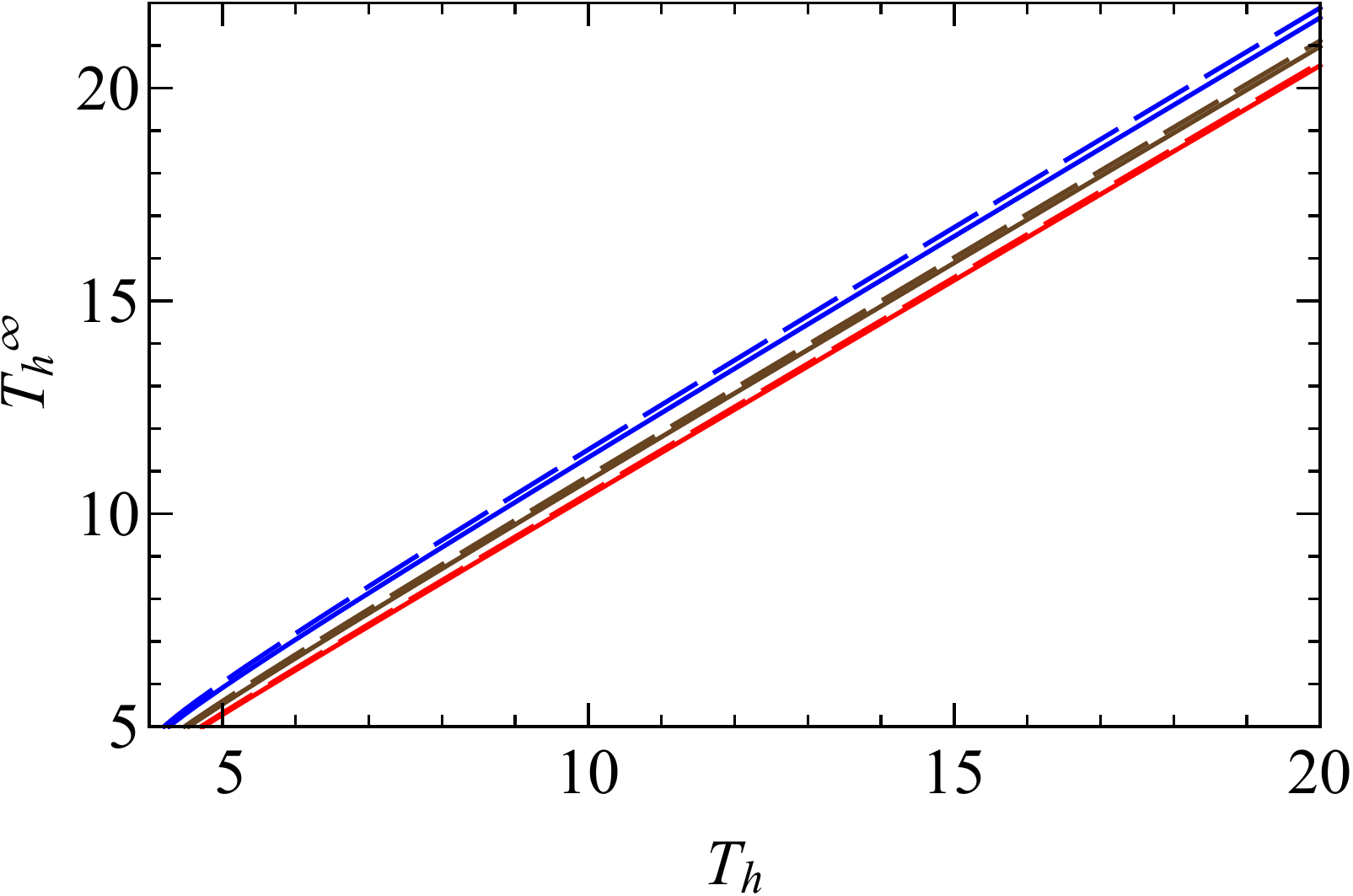}
\caption{(color online): $\Th^\infty$ as a function of observed $\Th$ for three bubble radii at $\TM$ for a saturated NaCl solution. From top to bottom: $r=0.6$, $1.2$, and $2.7\unit{\mu m}$. The solutions for the bubble spinodal and binodal are shown with solid and dashed curves, respectively. 
\label{fig:ThinfNaCl}
}
\end{figure}

These equations admit analytic solutions which are too lengthy to be displayed here. For relevant values of the parameters, three solutions are complex, and one real. We select the real solution for $R$ ($R_\mathrm{sp}$ from Eq.~\ref{eq:RsNaCl} and $R_\mathrm{eq}$ from Eq.~\ref{eq:ReNaCl}), and deduce the limits on $\Th^\infty$ from Eqs.~\ref{eq:ThsNaCl} and ~\ref{eq:TheNaCl}. The results are displayed in Fig.~\ref{fig:ThinfNaCl}. For easy comparison, we used the same three values of $r$ as in Fig.~\ref{fig:Thinf}; note that, because of the strong temperature dependence of $\rL^\infty$ for the brine, this corresponds to smaller inclusion volumes than if the same inclusion were filled with pure water.

\section{Fluid confined in a container partially wet by the liquid\label{sec:partial}}

In this section we turn to a fluid whose liquid phase wets the container walls only partially. This means that, when a vapor bubble is present, it will adopt a lenticular shape. We will assume that the contact angle of the liquid-vapor interface on the wall (measured on the liquid side) has a fixed value $\thc$, equal to its bulk value. We will need $\gamma$, $\gamma_\mathrm{LM}$, and $\gamma_\mathrm{VM}$, the interfacial tensions of the liquid-vapor, liquid-matrix, and vapor-matrix interfaces, respectively. They are connected with $\thc$ through the Young-Dupr\'{e} relation:
\beq
\gamma_\mathrm{VM} - \gamma_\mathrm{LM} = \gamma \cos \thc \, .
\label{eq:Young}
\eeq

\subsection{Modified free energy change in the canonical ensemble\label{sec:Dfpartial}}

The thermodynamic potential is still the Helmholtz free energy, but Eq.~\ref{eq:DF3} needs to be modified to account for the presence of new fluid-matrix interfaces:
\begin{eqnarray}
\Delta F = & \frac{\VV}{2 \kappa} \left( \frac{{\dav}^2}{1-x} -1 \right) + \gamma \SLV + (\gamma_\mathrm{VM} - \gamma_\mathrm{LM}) \SVM \nonumber \\
= & \frac{\VV}{2 \kappa} \left( \frac{{\dav}^2}{1-x} -1 \right) + \gamma ( \SLV + \SVM \cos \thc  ) \, ,
\label{eq:DFpartial1}
\end{eqnarray}
where $\SLV$ and $\SVM$ are the surface area of the liquid-vapor and vapor-matrix interfaces, respectively.

For simplicity we restrict our analysis to a smooth spherical container of radius $R$ and volume $V =(4/3) \pi R^3$. Note that more complex shapes can be encountered, and may affect the results, especially in the presence of microscopic, non-wetting defects such as grooves or pits~\cite{giacomello_geometry_2013}. Keeping the definition of $\phi= 2 \kappa \Delta F/V$, Eq.~\ref{eq:phi2} needs to be modified to account for the new geometry:
\beq
\phi = \left( \frac{{\dav}^2}{1-x} -1 \right) x + 3 \epsilon \sigma (x) \, ,
\label{eq:phipartial1}
\eeq
where the function $\sigma (x)$ quantifies the surface-to-volume ratio:
\beq
\sigma (x) = \frac{R ( \SLV + \SVM \cos \thc  )}{V} \, .
\label{eq:SVratio}
\eeq

\begin{figure}
\centering
\includegraphics[width=0.75\columnwidth]{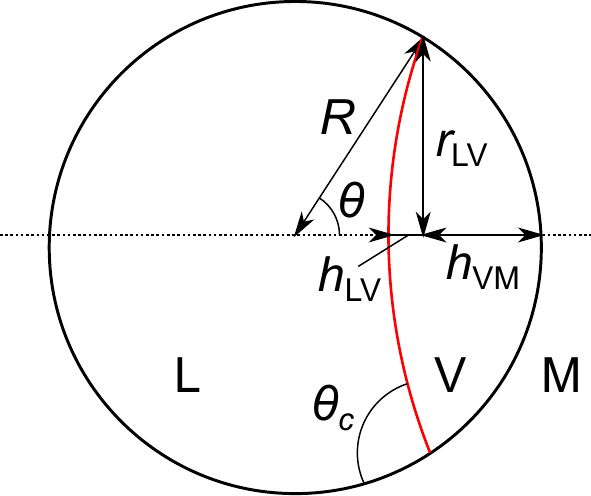}
\caption{(color online): Sketch of the system in the case of partial wetting with contact angle $\thc$. Letters L, V, and M indicate the liquid, vapor, and enclosing matrix, respectively.
\label{fig:sketch}
}
\end{figure}

In the case of complete wetting with a spherical bubble (Section~\ref{sec:wet}), $\sigma (x) = 3 x^{2/3}$. In the case of partial wetting, the calculation of $\sigma (x)$ is more involved. The system is sketched in Fig.~\ref{fig:sketch}, where we introduce the angle $\theta$ locating the liquid-vapor contact line, and various useful lengths. Trigonometry gives the following relations:
\begin{eqnarray}
\rLV(\theta)	&	=	&	R \, \sin \theta\\
\hLM(\theta)	&	=	&	R \, ( 1 - \cos \theta )\\
\hLV(\theta)	&	=	&	R \, \left[ 1 + \cos ( \thc + \theta ) \right] \frac{\sin \theta}{\sin ( \thc + \theta )}\\
\VL(\theta)	&	=	&	\frac{\pi}{6} \, \hLM (\theta) \left[ 3 \, \rLV (\theta)^2 + \hLM (\theta) ^2 \right]\nonumber\\
		&		&	+\frac{\pi}{6} \, \hLV (\theta) \left[ 3 \, \rLV (\theta)^2 + \hLV (\theta) ^2 \right]\\
\SLV(\theta)	&	=	&	\pi \left[ \rLV (\theta)^2 + \hLV (\theta) ^2 \right]\\
\SLM(\theta)	&	=	&	\pi \left[ \rLV (\theta)^2 + \hLM (\theta) ^2 \right]
\end{eqnarray}

Unfortunately there is no simple exact expression for $\phi(x)$. We note that $\sigma$ and $\phi$ are more easily written in terms of the variable $\theta$ rather than the reduced volume $x$. Combining the above equations yield a complex expression for $\phi (\theta)$. We shall give a simpler, approximate expression in the next section.

\subsection{Rescaled Berthelot-Laplace length in the small bubble approximation\label{sec:small}}

To obtain a simpler expression similar to Eq.~\ref{eq:phi2}, we assume that the bubble is sufficiently small, $x \ll 1$, for the angle $\theta$ which locates the contact line (Fig.~\ref{fig:sketch}) to be also small.

\begin{figure}
\centering
\includegraphics[width=0.95\columnwidth]{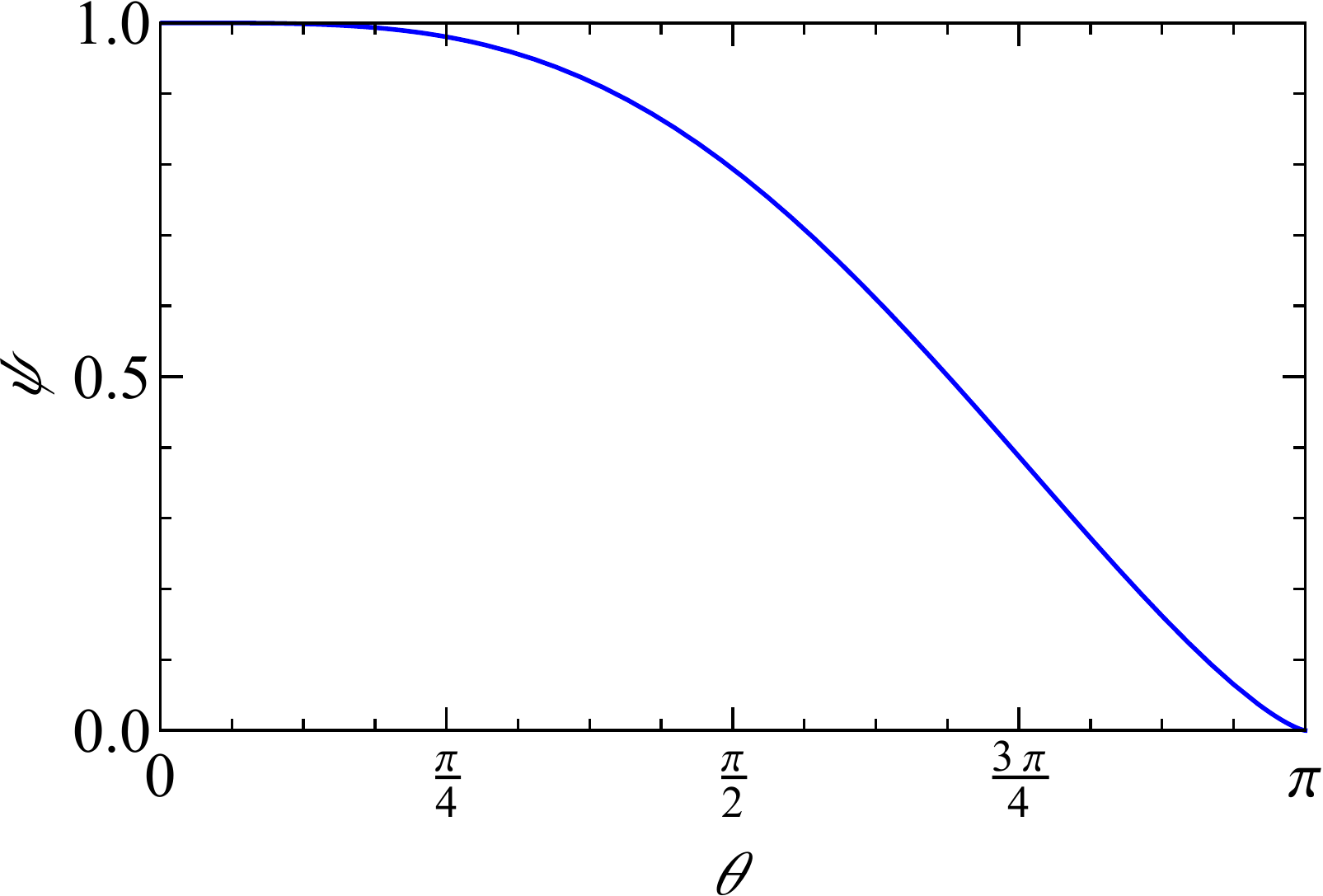}
\caption{(color online): Function $\psi$ (Eq.~\ref{eq:psi}) as a function of the contact angle $\thc$.
\label{fig:psi}
}
\end{figure}
We have:
\begin{eqnarray}
\label{eq:xsmallth} x(\theta) & = & \frac{(2-\cos \thc ) \cos \frac{\thc}{2}}{8\left( \sin \frac{\thc}{2} \right)^3} \,\theta^3 + \mathcal{O}(\theta^4) \, ,\\
\label{eq:ssmallth} \sigma(\theta) &=& \frac{3 \,(3 + 2 \cos \thc - \cos 2 \thc)}{16\left( \sin \frac{\thc}{2} \right)^2} \,\theta^2 + \mathcal{O}(\theta^3)\, .
\end{eqnarray}
To lowest order in $\theta$, we can write $\sigma = 3 \psi (\thc) x^{2/3}$, with:
\beq
\psi (\thc) = \frac{3 + 2 \cos \thc - \cos 2 \thc}{4\, \left[(2-\cos \thc ) \cos \frac{\thc}{2} \right]^{2/3}} \, .
\label{eq:psi}
\eeq
\begin{figure}[bbb]
\centering
\includegraphics[width=0.95\columnwidth]{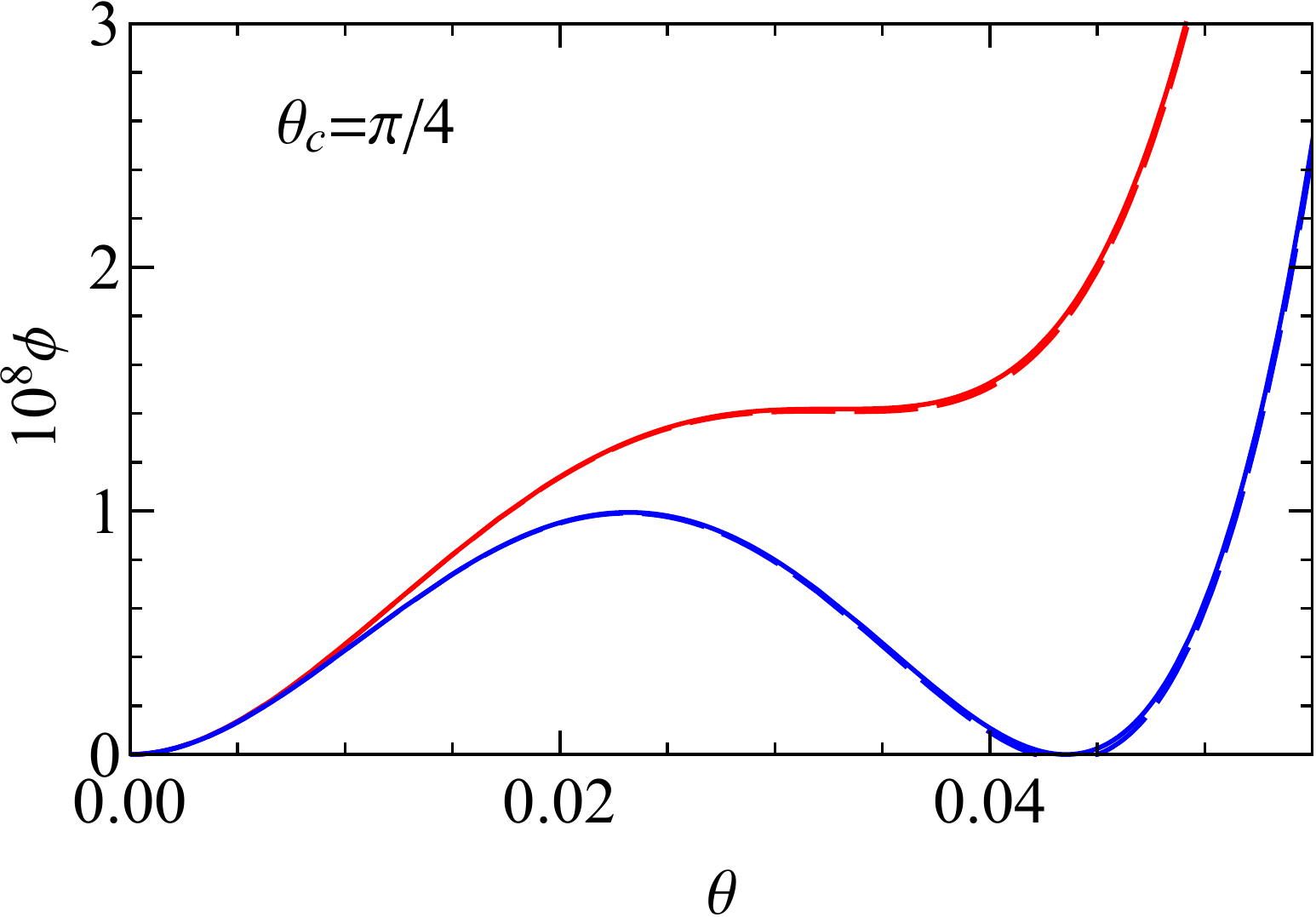}
\includegraphics[width=0.95\columnwidth]{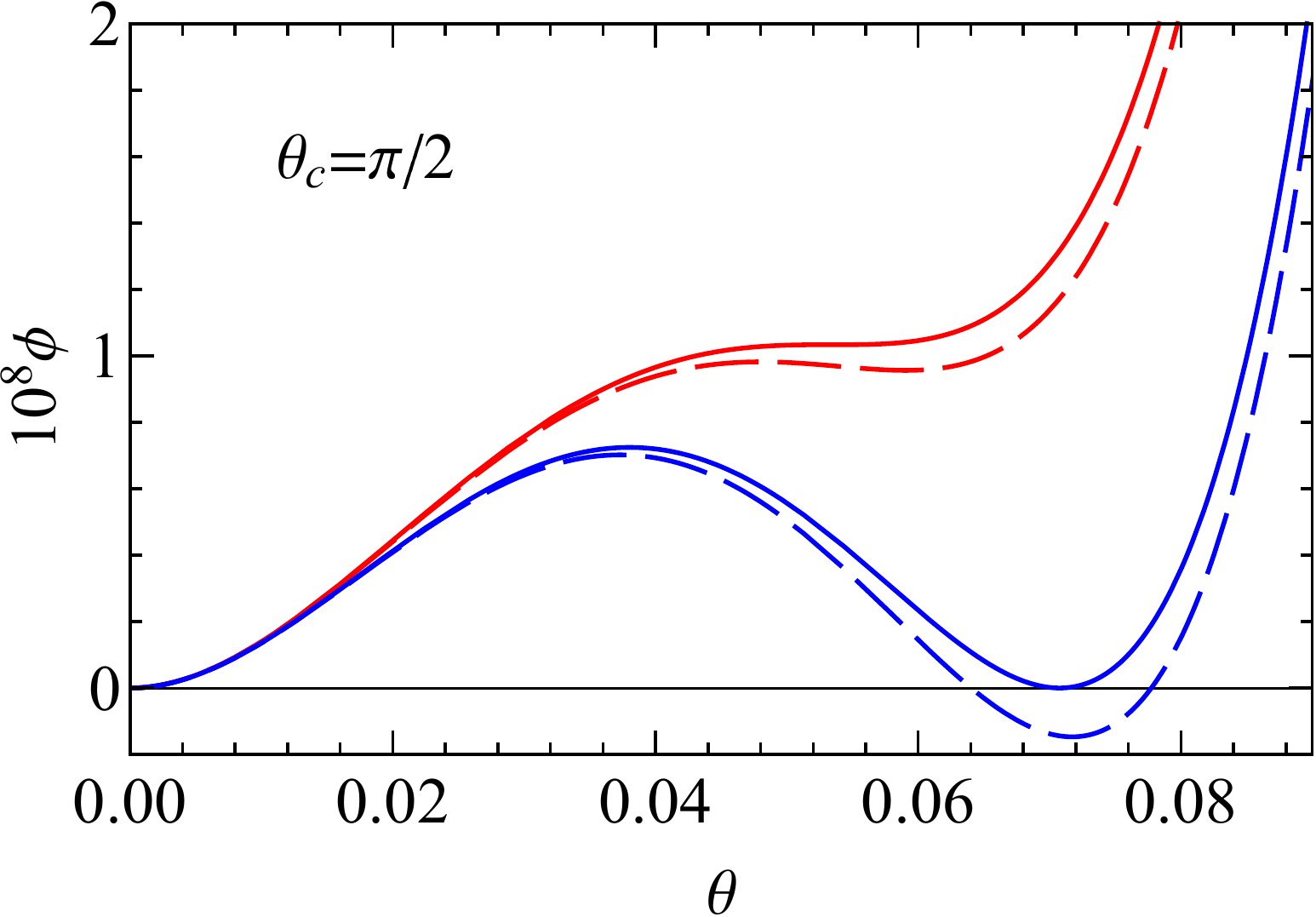}
\includegraphics[width=0.95\columnwidth]{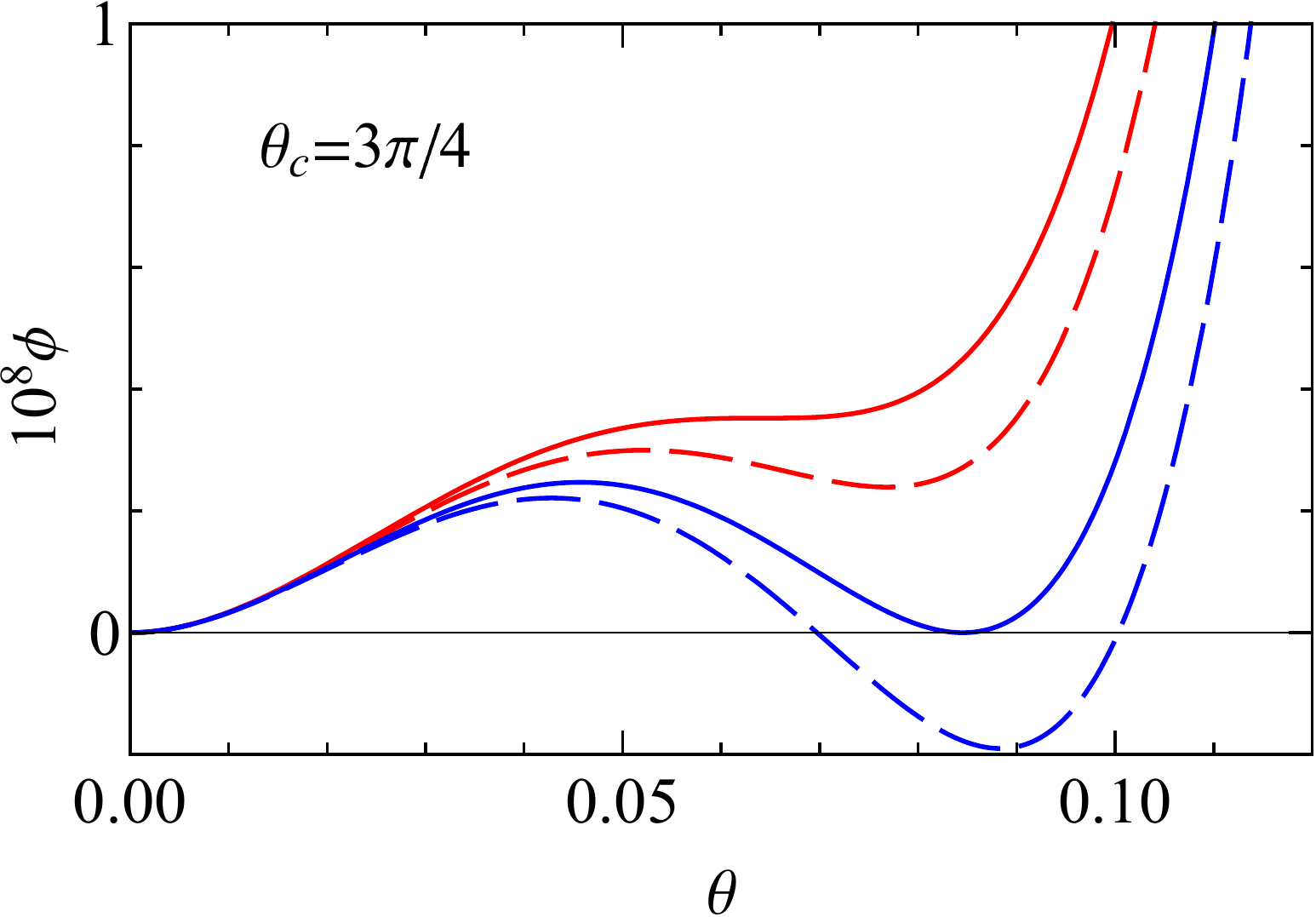}
\caption{(color online): Reduced free energy $\phi$ as a function of the angle $\theta$ locating the contact line for three values of the contact angle $\thc$ for pure water at $12\unit{\degreeC}$ and $V=10^3\,\mathrm{\mu ^3}$ ($\epsilon = 3.8\,10^{-6}$). The exact and approximate results are shown with solid and dashed curves, respectively, for the bubble spinodal (red) and binodal (blue).
\label{fig:phipart}
}
\end{figure}
We finally obtain:
\beq
\phi = \left( \frac{{\dav}^2}{1-x} -1 \right) x + 9 \epsilon \psi (\thc) x^{2/3} \, ,
\label{eq:phipartial2}
\eeq
As $\psi (0) = 1$, Eq.~\ref{eq:phi2} is recovered in the limit $\thc \rightarrow 0$ (complete wetting). For partial wetting, the only change to Eq.~\ref{eq:phi2} is that $\epsilon$ is replaced by $\epsilon \psi (\thc)$. We can therefore deduce from the results of Section~\ref{sec:wet} the modified results for the case of partial wetting by rescaling the Berthelot-Laplace length $\lambda$ (Section~\ref{sec:BL}) by a factor $\psi (\thc)$.

Figure~\ref{fig:psi} shows the function $\psi (\thc)$. It decreases significantly only at relatively large contact angles: at $\thc = 2 \pi/5$, $\psi$ is still 0.898. Therefore, the case of partial wetting with moderate contact angles will not differ much from the case of complete wetting.

Note however that the approximation $\theta \ll 1 $ fails for large contact angle $\thc$. The vapor will then adopt a film shape with a small $x$ but a large $\theta$. In the next section, we study a few cases, comparing the simple formulas in the small $\theta$ limit to the exact results.

\subsection{Results for various contact angles\label{sec:results}}

Figure~\ref{fig:phipart} provides the analog of Fig.~\ref{fig:phi} by displaying $\phi (\theta)$ for three choices of contact angles: $\thc = \pi/4$, $\pi/2$, and $3\pi/4$. We focus here on the bubble spinodal and binodal. We find $\ds$ and $\de$ from the exact function $\phi$ (Eq.~\ref{eq:phipartial1}), which is shown by solid curves. The small bubble approximation (Eq.~\ref{eq:phipartial2}) predicts $\ds$ and $\de$ by evaluating the complete wetting solutions (Section~\ref{sec:wet}) at $\epsilon = \psi(\thc) \lambda/R$. We show the exact function $\phi$ (Eq.~\ref{eq:phipartial1}) for these approximate values with dashed curves. As expected, the agreement is good for $\thc = \pi/4$ and deteriorates with increasing $\thc$. Still, the small bubble approximation provides a quick way to estimate $\ds$ and $\de$.

\section{Discussion and perspectives\label{sec:conclusion}}

Thanks to a generic fluid model, we now have the tools to correct the homogeneization temperature $\Th$ observed in fluid inclusions from the shift caused by surface tension, compressibility and finite volume. The analytic calculations give particularly simple expressions when making appropriate approximations. The approach is not limited to pure water, but covers any fluid with known surface tension and compressibility. This allowed us to show that the effect on $\Th$ in a saturated NaCl solution, which can be found included in evaporites, is significantly reduced compared to pure water, or to dilute solutions found in speleothems. As the Berthelot-Laplace length $\lambda$ varying only modestly over the full concentration range from pure water to brine, the origin of the reduced effect on $\Th$ stems from the steeper temperature dependence of the saturated liquid density in the brine. Indeed, when adding NaCl to water, the density maximum becomes less pronounced and shifted towards low temperatures, and eventually disappears above $1.5\unit{mol\,kg^{-1}}$.\cite{archer_thermodynamic_2000}

There is a workaround to avoid the limitations of the bubble-based method to determine the homogeneization temperature with microthermometry. It relies on Brillouin spectroscopy to measure the sound velocity as a function of temperature in the liquid, for the same fluid inclusion with and without a bubble. The intersection of these two curves directly gives $\Th^\infty$~\cite{mekki-azouzi_brillouin_2015}. The liquid pressure is still lower than $\Peq$ when the bubble is present, but this has a negligible effect on sound velocity. Using synthetic halite samples with known synthesis temperature, this technique was shown to be superior to microthermometry~\cite{guillerm_restoring_2020}, and it was applied to natural samples deposited in the Dead Sea basin during the last interglacial~\cite{brall_reconstructing_2022}.

More generally, confined fluids reveal subtle differences between thermodynamic ensembles~\cite{binder_van_2012}. When constrained to a fixed volume, the fluid shows interesting, and sometimes counter-intuitive, features. In particular, for small enough volumes, the confined liquid state remains stable at absolute negative pressure. This was noted before for fluid inclusions\cite{marti_effect_2012,wilhelmsen_communication_2014,wilhelmsen_evaluation_2015,vincent_statics_2017}. In fact, there are other occurrences of the same puzzling feature, even in open systems. Capillary condensation provides an example. Consider a material fully wet by the liquid, and a cylindrical pore of radius $R_p$ in this material. In the simplest approach~\cite{cohan_sorption_1938}, when the pore is exposed to a vapor pressure from $P^\infty$ to $P^\infty - 2\gamma/(\rL^\infty \kB T R)$, the stable state is a pore filled with liquid. The liquid-vapor interface adopts a spherical meniscus shape, whose radius decrease from infinity to $R$ when the vapor pressure decreases between the above limits. Consequently, the liquid pressure goes negative, down to $P^\infty - 2\gamma/(\rL^\infty \kB T R)$. Further pressure reduction empties the pore, with a liquid film left on the pore walls. More detailed treatments of the film are available~\cite{cole_excitation_1974,saam_excitations_1975}. A similar situation is encountered in the more open geometry of a sphere-plane contact. For a liquid that wets the surfaces (contact angle $\thc < \pi/2$), it is always favorable to replace the vapor by the liquid in the vicinity of the sphere-plane contact~\cite{caupin_comment_2008} and a liquid bridge forms spontaneously at any vapor pressure. Its shape can be fully calculated, and its mean radius of curvature $R_K$ is given by the Kelvin equation:
\beq
\frac{1}{R_K} = \frac{\rL^\infty \kb T}{\gamma} \log \frac{\PV}{P^\infty} \, .
\eeq
Here again, if $\PV$ is low enough, the liquid pressure becomes negative: $\PL = P^\infty + \rL^\infty \kb T \ln (\PV /\PL^\infty )= P^\infty - \gamma/{R_K}$.

Nevertheless, there is a difference between the cylindrical pore and the sphere-plane contact on the one hand, and the fluid inclusion in the other hand. In the former case, for a liquid that completely wets the walls, emptying occurs by recession of the liquid-vapor meniscus. It can then be argued that cavitation is prevented by the confinement: the critical radius for cavitation, $R_c = 2\gamma/(\PV - \PL)$, reaches the size of the container~\cite{caupin_comment_2008}; the system cannot accommodate the critical bubble and nucleation does not occur. In the latter case, Wilhelmsen~{\etal}~\cite{wilhelmsen_communication_2014} pointed out that the liquid state remains stable at negative pressure even in inclusions whose volume is larger than the critical radius. We confirm this finding: in our notations, at the bubble spinodal and for $\thc =0$, $R_c/R \simeq (3/4) \epsilon^{3/4} < 1$. Compressibility is required to explain the phenomenon of superstability.

At larger degrees of metastability (lower density at fixed $R$), the fluid becomes metastable and nucleation can occur. The confinement-compressibility effect modifies the energy barrier to nucleation. This was studied for complete wetting~\cite{wilhelmsen_evaluation_2015}. Our extension of the analysis to partial wetting opens the way to address more complex cases, such as cavitation in a small cubic box~\cite{pellegrin_cavitation_2020} or on nanodecorated surfaces~\cite{amabili_wetting_2016} with various wettability investigated by molecular dynamics simulations.

Of course, for the small degrees of metastability considered here, the metastable liquid state can also be observed easily, because the energy barrier for nucleation is much larger than the thermal energy~\cite{caupin_cavitation_2006}. If nucleation is homogeneous, one needs to reach pressures beyond $-100\unit{MPa}$ to observe cavitation\cite{zheng_limiting_2002,alvarenga_elastic_1993,shmulovich_experimental_2009,el_mekki_azouzi_coherent_2013,pallares_anomalies_2014,pallares_equation_2016,qiu_exploration_2016,holten_compressibility_2017}. Wilhelmsen~{\etal}~\cite{wilhelmsen_communication_2014} show that tiny containers, in the nanometer range, can achieve superstabilization at comparable pressures. However, when the dimensions are so close to the molecular sizes, the macroscopic thermodynamic treatment we have applied is questionable, and the properties of the fluid may differ from the bulk. It is therefore preferable to measure properties of liquids at negative pressure in metastable, micron-sized fluid inclusions. Still, the intriguing possibility to detect fluctuations between the homogeneous liquid and the bubble state with mesoscopic fluid inclusions (a couple of hundred nanometers, see Section~\ref{sec:eq}) deserves further investigations.

\acknowledgments
Support from Agence Nationale de la Recherche, grant number ANR-19-CE30-0035-01, is acknowledged.

%


\begin{thebibliography}{53}%
\makeatletter
\providecommand \@ifxundefined [1]{%
 \@ifx{#1\undefined}
}%
\providecommand \@ifnum [1]{%
 \ifnum #1\expandafter \@firstoftwo
 \else \expandafter \@secondoftwo
 \fi
}%
\providecommand \@ifx [1]{%
 \ifx #1\expandafter \@firstoftwo
 \else \expandafter \@secondoftwo
 \fi
}%
\providecommand \natexlab [1]{#1}%
\providecommand \enquote  [1]{``#1''}%
\providecommand \bibnamefont  [1]{#1}%
\providecommand \bibfnamefont [1]{#1}%
\providecommand \citenamefont [1]{#1}%
\providecommand \href@noop [0]{\@secondoftwo}%
\providecommand \href [0]{\begingroup \@sanitize@url \@href}%
\providecommand \@href[1]{\@@startlink{#1}\@@href}%
\providecommand \@@href[1]{\endgroup#1\@@endlink}%
\providecommand \@sanitize@url [0]{\catcode `\\12\catcode `\$12\catcode
  `\&12\catcode `\#12\catcode `\^12\catcode `\_12\catcode `\%12\relax}%
\providecommand \@@startlink[1]{}%
\providecommand \@@endlink[0]{}%
\providecommand \url  [0]{\begingroup\@sanitize@url \@url }%
\providecommand \@url [1]{\endgroup\@href {#1}{\urlprefix }}%
\providecommand \urlprefix  [0]{URL }%
\providecommand \Eprint [0]{\href }%
\providecommand \doibase [0]{https://doi.org/}%
\providecommand \selectlanguage [0]{\@gobble}%
\providecommand \bibinfo  [0]{\@secondoftwo}%
\providecommand \bibfield  [0]{\@secondoftwo}%
\providecommand \translation [1]{[#1]}%
\providecommand \BibitemOpen [0]{}%
\providecommand \bibitemStop [0]{}%
\providecommand \bibitemNoStop [0]{.\EOS\space}%
\providecommand \EOS [0]{\spacefactor3000\relax}%
\providecommand \BibitemShut  [1]{\csname bibitem#1\endcsname}%
\let\auto@bib@innerbib\@empty
\bibitem [{\citenamefont {Skinner}\ and\ \citenamefont
  {Sambles}(1972)}]{skinner_kelvin_1972}%
  \BibitemOpen
  \bibfield  {author} {\bibinfo {author} {\bibfnamefont {L.~M.}\ \bibnamefont
  {Skinner}}\ and\ \bibinfo {author} {\bibfnamefont {J.~R.}\ \bibnamefont
  {Sambles}},\ }\bibfield  {title} {\enquote {\bibinfo {title} {The {{Kelvin}}
  equation\textemdash a review},}\ }\href
  {https://doi.org/10.1016/0021-8502(72)90158-9} {\bibfield  {journal}
  {\bibinfo  {journal} {Journal of Aerosol Science}\ }\textbf {\bibinfo
  {volume} {3}},\ \bibinfo {pages} {199--210} (\bibinfo {year}
  {1972})}\BibitemShut {NoStop}%
\bibitem [{\citenamefont {Baletto}\ and\ \citenamefont
  {Ferrando}(2005)}]{baletto_structural_2005}%
  \BibitemOpen
  \bibfield  {author} {\bibinfo {author} {\bibfnamefont {F.}~\bibnamefont
  {Baletto}}\ and\ \bibinfo {author} {\bibfnamefont {R.}~\bibnamefont
  {Ferrando}},\ }\bibfield  {title} {\enquote {\bibinfo {title} {Structural
  properties of nanoclusters: Energetic, thermodynamic, and kinetic effects},}\
  }\href {https://doi.org/10.1103/RevModPhys.77.371} {\bibfield  {journal}
  {\bibinfo  {journal} {Rev. Mod. Phys.}\ }\textbf {\bibinfo {volume} {77}},\
  \bibinfo {pages} {371--423} (\bibinfo {year} {2005})}\BibitemShut {NoStop}%
\bibitem [{\citenamefont {Gelb}\ \emph {et~al.}(1999)\citenamefont {Gelb},
  \citenamefont {Gubbins}, \citenamefont {Radhakrishnan},\ and\ \citenamefont
  {{Sliwinska-Bartkowiak}}}]{gelb_phase_1999}%
  \BibitemOpen
  \bibfield  {author} {\bibinfo {author} {\bibfnamefont {L.~D.}\ \bibnamefont
  {Gelb}}, \bibinfo {author} {\bibfnamefont {K.~E.}\ \bibnamefont {Gubbins}},
  \bibinfo {author} {\bibfnamefont {R.}~\bibnamefont {Radhakrishnan}},\ and\
  \bibinfo {author} {\bibfnamefont {M.}~\bibnamefont
  {{Sliwinska-Bartkowiak}}},\ }\bibfield  {title} {\enquote {\bibinfo {title}
  {Phase separation in confined systems},}\ }\href
  {https://doi.org/10.1088/0034-4885/62/12/201} {\bibfield  {journal} {\bibinfo
   {journal} {Rep. Prog. Phys.}\ }\textbf {\bibinfo {volume} {62}},\ \bibinfo
  {pages} {1573} (\bibinfo {year} {1999})}\BibitemShut {NoStop}%
\bibitem [{\citenamefont {Huber}(2015)}]{huber_soft_2015}%
  \BibitemOpen
  \bibfield  {author} {\bibinfo {author} {\bibfnamefont {P.}~\bibnamefont
  {Huber}},\ }\bibfield  {title} {\enquote {\bibinfo {title} {Soft matter in
  hard confinement: Phase transition thermodynamics, structure, texture,
  diffusion and flow in nanoporous media},}\ }\href@noop {} {\bibfield
  {journal} {\bibinfo  {journal} {J. Phys.}\ ,\ \bibinfo {pages} {44}}
  (\bibinfo {year} {2015})}\BibitemShut {NoStop}%
\bibitem [{\citenamefont {{Alba-Simionesco}}\ \emph {et~al.}(2006)\citenamefont
  {{Alba-Simionesco}}, \citenamefont {Coasne}, \citenamefont {Dosseh},
  \citenamefont {Dudziak}, \citenamefont {Gubbins}, \citenamefont
  {Radhakrishnan},\ and\ \citenamefont
  {{Sliwinska-Bartkowiak}}}]{alba-simionesco_effects_2006}%
  \BibitemOpen
  \bibfield  {author} {\bibinfo {author} {\bibfnamefont {C.}~\bibnamefont
  {{Alba-Simionesco}}}, \bibinfo {author} {\bibfnamefont {B.}~\bibnamefont
  {Coasne}}, \bibinfo {author} {\bibfnamefont {G.}~\bibnamefont {Dosseh}},
  \bibinfo {author} {\bibfnamefont {G.}~\bibnamefont {Dudziak}}, \bibinfo
  {author} {\bibfnamefont {K.~E.}\ \bibnamefont {Gubbins}}, \bibinfo {author}
  {\bibfnamefont {R.}~\bibnamefont {Radhakrishnan}},\ and\ \bibinfo {author}
  {\bibfnamefont {M.}~\bibnamefont {{Sliwinska-Bartkowiak}}},\ }\bibfield
  {title} {\enquote {\bibinfo {title} {Effects of confinement on freezing and
  melting},}\ }\href {https://doi.org/10.1088/0953-8984/18/6/R01} {\bibfield
  {journal} {\bibinfo  {journal} {J. Phys. Condens. Matter}\ }\textbf {\bibinfo
  {volume} {18}},\ \bibinfo {pages} {R15--R68} (\bibinfo {year}
  {2006})}\BibitemShut {NoStop}%
\bibitem [{\citenamefont {Xu}\ \emph {et~al.}(2006)\citenamefont {Xu},
  \citenamefont {Sharp}, \citenamefont {Yuan}, \citenamefont {Yi},
  \citenamefont {Liao}, \citenamefont {Glaeser}, \citenamefont {Minor},
  \citenamefont {Beeman}, \citenamefont {Ridgway}, \citenamefont {Kluth},
  \citenamefont {Ager}, \citenamefont {Chrzan},\ and\ \citenamefont
  {Haller}}]{xu_large_2006}%
  \BibitemOpen
  \bibfield  {author} {\bibinfo {author} {\bibfnamefont {Q.}~\bibnamefont
  {Xu}}, \bibinfo {author} {\bibfnamefont {I.}~\bibnamefont {Sharp}}, \bibinfo
  {author} {\bibfnamefont {C.}~\bibnamefont {Yuan}}, \bibinfo {author}
  {\bibfnamefont {D.}~\bibnamefont {Yi}}, \bibinfo {author} {\bibfnamefont
  {C.}~\bibnamefont {Liao}}, \bibinfo {author} {\bibfnamefont {A.}~\bibnamefont
  {Glaeser}}, \bibinfo {author} {\bibfnamefont {A.}~\bibnamefont {Minor}},
  \bibinfo {author} {\bibfnamefont {J.}~\bibnamefont {Beeman}}, \bibinfo
  {author} {\bibfnamefont {M.}~\bibnamefont {Ridgway}}, \bibinfo {author}
  {\bibfnamefont {P.}~\bibnamefont {Kluth}}, \bibinfo {author} {\bibfnamefont
  {J.}~\bibnamefont {Ager}}, \bibinfo {author} {\bibfnamefont {D.}~\bibnamefont
  {Chrzan}},\ and\ \bibinfo {author} {\bibfnamefont {E.}~\bibnamefont
  {Haller}},\ }\bibfield  {title} {\enquote {\bibinfo {title} {Large
  melting-point hysteresis of {{Ge}} nanocrystals embedded in
  {{SiO}}{\textsubscript{2}}},}\ }\href
  {https://doi.org/10.1103/PhysRevLett.97.155701} {\bibfield  {journal}
  {\bibinfo  {journal} {Phys. Rev. Lett.}\ }\textbf {\bibinfo {volume} {97}},\
  \bibinfo {pages} {155701} (\bibinfo {year} {2006})}\BibitemShut {NoStop}%
\bibitem [{\citenamefont {Caupin}(2008)}]{caupin_melting_2008}%
  \BibitemOpen
  \bibfield  {author} {\bibinfo {author} {\bibfnamefont {F.}~\bibnamefont
  {Caupin}},\ }\bibfield  {title} {\enquote {\bibinfo {title} {Melting and
  freezing of embedded nanoclusters},}\ }\href
  {https://doi.org/10.1103/PhysRevB.77.184108} {\bibfield  {journal} {\bibinfo
  {journal} {Phys. Rev. B}\ }\textbf {\bibinfo {volume} {77}},\ \bibinfo
  {pages} {184108} (\bibinfo {year} {2008})}\BibitemShut {NoStop}%
\bibitem [{\citenamefont {Caupin}(2007)}]{caupin_comment_2007a}%
  \BibitemOpen
  \bibfield  {author} {\bibinfo {author} {\bibfnamefont {F.}~\bibnamefont
  {Caupin}},\ }\bibfield  {title} {\enquote {\bibinfo {title} {Comment on
  ``{{Large}} melting-point hysteresis of {{Ge}} nanocrystals embedded in
  {{SiO}}{\textsubscript{2}}''},}\ }\href
  {https://doi.org/10.1103/PhysRevLett.99.079601} {\bibfield  {journal}
  {\bibinfo  {journal} {Phys. Rev. Lett.}\ }\textbf {\bibinfo {volume} {99}},\
  \bibinfo {pages} {079601} (\bibinfo {year} {2007})}\BibitemShut {NoStop}%
\bibitem [{\citenamefont {Xu}\ \emph {et~al.}(2007)\citenamefont {Xu},
  \citenamefont {Sharp}, \citenamefont {Yuan}, \citenamefont {Yi},
  \citenamefont {Liao}, \citenamefont {Glaeser}, \citenamefont {Minor},
  \citenamefont {Beeman}, \citenamefont {Ridgway}, \citenamefont {Kluth},
  \citenamefont {Ager}, \citenamefont {Chrzan},\ and\ \citenamefont
  {Haller}}]{xu_xu_2007}%
  \BibitemOpen
  \bibfield  {author} {\bibinfo {author} {\bibfnamefont {Q.}~\bibnamefont
  {Xu}}, \bibinfo {author} {\bibfnamefont {I.~D.}\ \bibnamefont {Sharp}},
  \bibinfo {author} {\bibfnamefont {C.~W.}\ \bibnamefont {Yuan}}, \bibinfo
  {author} {\bibfnamefont {D.~O.}\ \bibnamefont {Yi}}, \bibinfo {author}
  {\bibfnamefont {C.~Y.}\ \bibnamefont {Liao}}, \bibinfo {author}
  {\bibfnamefont {A.~M.}\ \bibnamefont {Glaeser}}, \bibinfo {author}
  {\bibfnamefont {A.~M.}\ \bibnamefont {Minor}}, \bibinfo {author}
  {\bibfnamefont {J.~W.}\ \bibnamefont {Beeman}}, \bibinfo {author}
  {\bibfnamefont {M.~C.}\ \bibnamefont {Ridgway}}, \bibinfo {author}
  {\bibfnamefont {P.}~\bibnamefont {Kluth}}, \bibinfo {author} {\bibfnamefont
  {J.~W.}\ \bibnamefont {Ager}}, \bibinfo {author} {\bibfnamefont {D.~C.}\
  \bibnamefont {Chrzan}},\ and\ \bibinfo {author} {\bibfnamefont {E.~E.}\
  \bibnamefont {Haller}},\ }\bibfield  {title} {\enquote {\bibinfo {title} {Xu
  {\emph{et al.}} reply:},}\ }\href
  {https://doi.org/10.1103/PhysRevLett.99.079602} {\bibfield  {journal}
  {\bibinfo  {journal} {Phys. Rev. Lett.}\ }\textbf {\bibinfo {volume} {99}},\
  \bibinfo {pages} {079602} (\bibinfo {year} {2007})}\BibitemShut {NoStop}%
\bibitem [{\citenamefont {Berthelot}(1850)}]{berthelot_sur_1850}%
  \BibitemOpen
  \bibfield  {author} {\bibinfo {author} {\bibfnamefont {M.}~\bibnamefont
  {Berthelot}},\ }\bibfield  {title} {\enquote {\bibinfo {title} {Sur quelques
  ph\'enom\`enes de dilatation forc\'ee des liquides},}\ }\href@noop {}
  {\bibfield  {journal} {\bibinfo  {journal} {Ann. Chim. Phys.}\ }\textbf
  {\bibinfo {volume} {30}},\ \bibinfo {pages} {232--237} (\bibinfo {year}
  {1850})}\BibitemShut {NoStop}%
\bibitem [{\citenamefont {Caupin}\ and\ \citenamefont
  {Herbert}(2006)}]{caupin_cavitation_2006}%
  \BibitemOpen
  \bibfield  {author} {\bibinfo {author} {\bibfnamefont {F.}~\bibnamefont
  {Caupin}}\ and\ \bibinfo {author} {\bibfnamefont {E.}~\bibnamefont
  {Herbert}},\ }\bibfield  {title} {\enquote {\bibinfo {title} {Cavitation in
  water: A review},}\ }\href {https://doi.org/10.1016/j.crhy.2006.10.015}
  {\bibfield  {journal} {\bibinfo  {journal} {Comptes Rendus Phys.}\ }\textbf
  {\bibinfo {volume} {7}},\ \bibinfo {pages} {1000--1017} (\bibinfo {year}
  {2006})}\BibitemShut {NoStop}%
\bibitem [{\citenamefont {Roedder}(1967)}]{roedder_metastable_1967}%
  \BibitemOpen
  \bibfield  {author} {\bibinfo {author} {\bibfnamefont {E.}~\bibnamefont
  {Roedder}},\ }\bibfield  {title} {\enquote {\bibinfo {title} {Metastable
  superheated ice in liquid-water inclusions under high negative pressure},}\
  }\href {https://doi.org/10.1126/science.155.3768.1413} {\bibfield  {journal}
  {\bibinfo  {journal} {Science}\ }\textbf {\bibinfo {volume} {155}},\ \bibinfo
  {pages} {1413--1417} (\bibinfo {year} {1967})}\BibitemShut {NoStop}%
\bibitem [{\citenamefont {Zheng}\ \emph {et~al.}(2002)\citenamefont {Zheng},
  \citenamefont {Green}, \citenamefont {Kieffer}, \citenamefont {Poole},
  \citenamefont {Shao}, \citenamefont {Wolf},\ and\ \citenamefont
  {Austen~Angell}}]{zheng_limiting_2002}%
  \BibitemOpen
  \bibfield  {author} {\bibinfo {author} {\bibfnamefont {Q.}~\bibnamefont
  {Zheng}}, \bibinfo {author} {\bibfnamefont {J.}~\bibnamefont {Green}},
  \bibinfo {author} {\bibfnamefont {J.}~\bibnamefont {Kieffer}}, \bibinfo
  {author} {\bibfnamefont {P.~H.}\ \bibnamefont {Poole}}, \bibinfo {author}
  {\bibfnamefont {J.}~\bibnamefont {Shao}}, \bibinfo {author} {\bibfnamefont
  {G.~H.}\ \bibnamefont {Wolf}},\ and\ \bibinfo {author} {\bibfnamefont
  {C.}~\bibnamefont {Austen~Angell}},\ }\bibfield  {title} {\enquote {\bibinfo
  {title} {Limiting tensions for liquids and glasses from laboratory and {{MD}}
  studies},}\ }in\ \href
  {http://www.springerlink.com/content/t4r2725487423504/abstract/} {\emph
  {\bibinfo {booktitle} {Liquids {{Under Negative Pressure}}}}},\ Vol.~\bibinfo
  {volume} {84},\ \bibinfo {editor} {edited by\ \bibinfo {editor}
  {\bibfnamefont {A.~R.}\ \bibnamefont {Imre}}, \bibinfo {editor}
  {\bibfnamefont {H.~J.}\ \bibnamefont {Maris}},\ and\ \bibinfo {editor}
  {\bibfnamefont {P.~R.}\ \bibnamefont {Williams}}}\ (\bibinfo  {publisher}
  {{Springer Netherlands}},\ \bibinfo {year} {2002})\ pp.\ \bibinfo {pages}
  {33--46}\BibitemShut {NoStop}%
\bibitem [{\citenamefont {Alvarenga}, \citenamefont {Grimsditch},\ and\
  \citenamefont {Bodnar}(1993)}]{alvarenga_elastic_1993}%
  \BibitemOpen
  \bibfield  {author} {\bibinfo {author} {\bibfnamefont {A.~D.}\ \bibnamefont
  {Alvarenga}}, \bibinfo {author} {\bibfnamefont {M.}~\bibnamefont
  {Grimsditch}},\ and\ \bibinfo {author} {\bibfnamefont {R.~J.}\ \bibnamefont
  {Bodnar}},\ }\bibfield  {title} {\enquote {\bibinfo {title} {Elastic
  properties of water under negative pressures},}\ }\href
  {https://doi.org/10.1063/1.464497} {\bibfield  {journal} {\bibinfo  {journal}
  {J. Chem. Phys.}\ }\textbf {\bibinfo {volume} {98}},\ \bibinfo {pages}
  {8392--8396} (\bibinfo {year} {1993})}\BibitemShut {NoStop}%
\bibitem [{\citenamefont {Shmulovich}\ \emph {et~al.}(2009)\citenamefont
  {Shmulovich}, \citenamefont {Mercury}, \citenamefont {Thi{\'e}ry},
  \citenamefont {Ramboz},\ and\ \citenamefont
  {El~Mekki}}]{shmulovich_experimental_2009}%
  \BibitemOpen
  \bibfield  {author} {\bibinfo {author} {\bibfnamefont {K.~I.}\ \bibnamefont
  {Shmulovich}}, \bibinfo {author} {\bibfnamefont {L.}~\bibnamefont {Mercury}},
  \bibinfo {author} {\bibfnamefont {R.}~\bibnamefont {Thi{\'e}ry}}, \bibinfo
  {author} {\bibfnamefont {C.}~\bibnamefont {Ramboz}},\ and\ \bibinfo {author}
  {\bibfnamefont {M.}~\bibnamefont {El~Mekki}},\ }\bibfield  {title} {\enquote
  {\bibinfo {title} {Experimental superheating of water and aqueous
  solutions},}\ }\href {https://doi.org/10.1016/j.gca.2009.02.006} {\bibfield
  {journal} {\bibinfo  {journal} {Geochim. Cosmochim. Acta}\ }\textbf {\bibinfo
  {volume} {73}},\ \bibinfo {pages} {2457--2470} (\bibinfo {year}
  {2009})}\BibitemShut {NoStop}%
\bibitem [{\citenamefont {El~Mekki~Azouzi}\ \emph {et~al.}(2013)\citenamefont
  {El~Mekki~Azouzi}, \citenamefont {Ramboz}, \citenamefont {Lenain},\ and\
  \citenamefont {Caupin}}]{el_mekki_azouzi_coherent_2013}%
  \BibitemOpen
  \bibfield  {author} {\bibinfo {author} {\bibfnamefont {M.}~\bibnamefont
  {El~Mekki~Azouzi}}, \bibinfo {author} {\bibfnamefont {C.}~\bibnamefont
  {Ramboz}}, \bibinfo {author} {\bibfnamefont {J.-F.}\ \bibnamefont {Lenain}},\
  and\ \bibinfo {author} {\bibfnamefont {F.}~\bibnamefont {Caupin}},\
  }\bibfield  {title} {\enquote {\bibinfo {title} {A coherent picture of water
  at extreme negative pressure},}\ }\href {https://doi.org/10.1038/nphys2475}
  {\bibfield  {journal} {\bibinfo  {journal} {Nat. Phys.}\ }\textbf {\bibinfo
  {volume} {9}},\ \bibinfo {pages} {38--41} (\bibinfo {year}
  {2013})}\BibitemShut {NoStop}%
\bibitem [{\citenamefont {Pallares}\ \emph {et~al.}(2014)\citenamefont
  {Pallares}, \citenamefont {Azouzi}, \citenamefont {Gonz{\'a}lez},
  \citenamefont {Aragones}, \citenamefont {Abascal}, \citenamefont
  {Valeriani},\ and\ \citenamefont {Caupin}}]{pallares_anomalies_2014}%
  \BibitemOpen
  \bibfield  {author} {\bibinfo {author} {\bibfnamefont {G.}~\bibnamefont
  {Pallares}}, \bibinfo {author} {\bibfnamefont {M.~E.~M.}\ \bibnamefont
  {Azouzi}}, \bibinfo {author} {\bibfnamefont {M.~A.}\ \bibnamefont
  {Gonz{\'a}lez}}, \bibinfo {author} {\bibfnamefont {J.~L.}\ \bibnamefont
  {Aragones}}, \bibinfo {author} {\bibfnamefont {J.~L.~F.}\ \bibnamefont
  {Abascal}}, \bibinfo {author} {\bibfnamefont {C.}~\bibnamefont {Valeriani}},\
  and\ \bibinfo {author} {\bibfnamefont {F.}~\bibnamefont {Caupin}},\
  }\bibfield  {title} {\enquote {\bibinfo {title} {Anomalies in bulk
  supercooled water at negative pressure},}\ }\href
  {https://doi.org/10.1073/pnas.1323366111} {\bibfield  {journal} {\bibinfo
  {journal} {Proc. Natl. Acad. Sci. USA}\ }\textbf {\bibinfo {volume} {111}},\
  \bibinfo {pages} {7936--7941} (\bibinfo {year} {2014})}\BibitemShut {NoStop}%
\bibitem [{\citenamefont {Pallares}\ \emph {et~al.}(2016)\citenamefont
  {Pallares}, \citenamefont {Gonzalez}, \citenamefont {Abascal}, \citenamefont
  {Valeriani},\ and\ \citenamefont {Caupin}}]{pallares_equation_2016}%
  \BibitemOpen
  \bibfield  {author} {\bibinfo {author} {\bibfnamefont {G.}~\bibnamefont
  {Pallares}}, \bibinfo {author} {\bibfnamefont {M.~A.}\ \bibnamefont
  {Gonzalez}}, \bibinfo {author} {\bibfnamefont {J.~L.~F.}\ \bibnamefont
  {Abascal}}, \bibinfo {author} {\bibfnamefont {C.}~\bibnamefont {Valeriani}},\
  and\ \bibinfo {author} {\bibfnamefont {F.}~\bibnamefont {Caupin}},\
  }\bibfield  {title} {\enquote {\bibinfo {title} {Equation of state for water
  and its line of density maxima down to -120 {{MPa}}},}\ }\href
  {https://doi.org/10.1039/C5CP07580G} {\bibfield  {journal} {\bibinfo
  {journal} {Phys. Chem. Chem. Phys.}\ }\textbf {\bibinfo {volume} {18}},\
  \bibinfo {pages} {5896--5900} (\bibinfo {year} {2016})}\BibitemShut {NoStop}%
\bibitem [{\citenamefont {Qiu}\ \emph {et~al.}(2016)\citenamefont {Qiu},
  \citenamefont {Kr{\"u}ger}, \citenamefont {Wilke}, \citenamefont {Marti},
  \citenamefont {Ri{\v c}ka},\ and\ \citenamefont
  {Frenz}}]{qiu_exploration_2016}%
  \BibitemOpen
  \bibfield  {author} {\bibinfo {author} {\bibfnamefont {C.}~\bibnamefont
  {Qiu}}, \bibinfo {author} {\bibfnamefont {Y.}~\bibnamefont {Kr{\"u}ger}},
  \bibinfo {author} {\bibfnamefont {M.}~\bibnamefont {Wilke}}, \bibinfo
  {author} {\bibfnamefont {D.}~\bibnamefont {Marti}}, \bibinfo {author}
  {\bibfnamefont {J.}~\bibnamefont {Ri{\v c}ka}},\ and\ \bibinfo {author}
  {\bibfnamefont {M.}~\bibnamefont {Frenz}},\ }\bibfield  {title} {\enquote
  {\bibinfo {title} {Exploration of the phase diagram of liquid water in the
  low-temperature metastable region using synthetic fluid inclusions},}\ }\href
  {https://doi.org/10.1039/C6CP04250C} {\bibfield  {journal} {\bibinfo
  {journal} {Phys. Chem. Chem. Phys.}\ }\textbf {\bibinfo {volume} {18}},\
  \bibinfo {pages} {28227--28241} (\bibinfo {year} {2016})}\BibitemShut
  {NoStop}%
\bibitem [{\citenamefont {Holten}\ \emph {et~al.}(2017)\citenamefont {Holten},
  \citenamefont {Qiu}, \citenamefont {Guillerm}, \citenamefont {Wilke},
  \citenamefont {Ri{\v c}ka}, \citenamefont {Frenz},\ and\ \citenamefont
  {Caupin}}]{holten_compressibility_2017}%
  \BibitemOpen
  \bibfield  {author} {\bibinfo {author} {\bibfnamefont {V.}~\bibnamefont
  {Holten}}, \bibinfo {author} {\bibfnamefont {C.}~\bibnamefont {Qiu}},
  \bibinfo {author} {\bibfnamefont {E.}~\bibnamefont {Guillerm}}, \bibinfo
  {author} {\bibfnamefont {M.}~\bibnamefont {Wilke}}, \bibinfo {author}
  {\bibfnamefont {J.}~\bibnamefont {Ri{\v c}ka}}, \bibinfo {author}
  {\bibfnamefont {M.}~\bibnamefont {Frenz}},\ and\ \bibinfo {author}
  {\bibfnamefont {F.}~\bibnamefont {Caupin}},\ }\bibfield  {title} {\enquote
  {\bibinfo {title} {Compressibility anomalies in stretched water and their
  interplay with density anomalies},}\ }\href
  {https://doi.org/10.1021/acs.jpclett.7b02563} {\bibfield  {journal} {\bibinfo
   {journal} {J. Phys. Chem. Lett.}\ }\textbf {\bibinfo {volume} {8}},\
  \bibinfo {pages} {5519--5522} (\bibinfo {year} {2017})}\BibitemShut {NoStop}%
\bibitem [{\citenamefont {Marti}\ \emph {et~al.}(2012)\citenamefont {Marti},
  \citenamefont {Kr{\"u}ger}, \citenamefont {Fleitmann}, \citenamefont
  {Frenz},\ and\ \citenamefont {Ri{\v c}ka}}]{marti_effect_2012}%
  \BibitemOpen
  \bibfield  {author} {\bibinfo {author} {\bibfnamefont {D.}~\bibnamefont
  {Marti}}, \bibinfo {author} {\bibfnamefont {Y.}~\bibnamefont {Kr{\"u}ger}},
  \bibinfo {author} {\bibfnamefont {D.}~\bibnamefont {Fleitmann}}, \bibinfo
  {author} {\bibfnamefont {M.}~\bibnamefont {Frenz}},\ and\ \bibinfo {author}
  {\bibfnamefont {J.}~\bibnamefont {Ri{\v c}ka}},\ }\bibfield  {title}
  {\enquote {\bibinfo {title} {The effect of surface tension on
  liquid\textendash gas equilibria in isochoric systems and its application to
  fluid inclusions},}\ }\href {https://doi.org/10.1016/j.fluid.2011.08.010}
  {\bibfield  {journal} {\bibinfo  {journal} {Fluid Phase Equilib.}\ }\textbf
  {\bibinfo {volume} {314}},\ \bibinfo {pages} {13--21} (\bibinfo {year}
  {2012})}\BibitemShut {NoStop}%
\bibitem [{\citenamefont {Wilhelmsen}\ \emph {et~al.}(2014)\citenamefont
  {Wilhelmsen}, \citenamefont {Bedeaux}, \citenamefont {Kjelstrup},\ and\
  \citenamefont {Reguera}}]{wilhelmsen_communication_2014}%
  \BibitemOpen
  \bibfield  {author} {\bibinfo {author} {\bibfnamefont {{\O}.}~\bibnamefont
  {Wilhelmsen}}, \bibinfo {author} {\bibfnamefont {D.}~\bibnamefont {Bedeaux}},
  \bibinfo {author} {\bibfnamefont {S.}~\bibnamefont {Kjelstrup}},\ and\
  \bibinfo {author} {\bibfnamefont {D.}~\bibnamefont {Reguera}},\ }\bibfield
  {title} {\enquote {\bibinfo {title} {Communication: Superstabilization of
  fluids in nanocontainers},}\ }\href {https://doi.org/10.1063/1.4893701}
  {\bibfield  {journal} {\bibinfo  {journal} {The Journal of Chemical Physics}\
  }\textbf {\bibinfo {volume} {141}},\ \bibinfo {pages} {071103} (\bibinfo
  {year} {2014})}\BibitemShut {NoStop}%
\bibitem [{\citenamefont {Wilhelmsen}\ and\ \citenamefont
  {Reguera}(2015)}]{wilhelmsen_evaluation_2015}%
  \BibitemOpen
  \bibfield  {author} {\bibinfo {author} {\bibfnamefont {{\O}.}~\bibnamefont
  {Wilhelmsen}}\ and\ \bibinfo {author} {\bibfnamefont {D.}~\bibnamefont
  {Reguera}},\ }\bibfield  {title} {\enquote {\bibinfo {title} {Evaluation of
  finite-size effects in cavitation and droplet formation},}\ }\href
  {https://doi.org/10.1063/1.4907367} {\bibfield  {journal} {\bibinfo
  {journal} {The Journal of Chemical Physics}\ }\textbf {\bibinfo {volume}
  {142}},\ \bibinfo {pages} {064703} (\bibinfo {year} {2015})}\BibitemShut
  {NoStop}%
\bibitem [{\citenamefont {Vincent}\ and\ \citenamefont
  {Marmottant}(2017)}]{vincent_statics_2017}%
  \BibitemOpen
  \bibfield  {author} {\bibinfo {author} {\bibfnamefont {O.}~\bibnamefont
  {Vincent}}\ and\ \bibinfo {author} {\bibfnamefont {P.}~\bibnamefont
  {Marmottant}},\ }\bibfield  {title} {\enquote {\bibinfo {title} {On the
  statics and dynamics of fully confined bubbles},}\ }\href
  {https://doi.org/10.1017/jfm.2017.487} {\bibfield  {journal} {\bibinfo
  {journal} {J. Fluid Mech.}\ }\textbf {\bibinfo {volume} {827}},\ \bibinfo
  {pages} {194--224} (\bibinfo {year} {2017})}\BibitemShut {NoStop}%
\bibitem [{\citenamefont {Fall}, \citenamefont {Rimstidt},\ and\ \citenamefont
  {Bodnar}(2009)}]{fall_effect_2009}%
  \BibitemOpen
  \bibfield  {author} {\bibinfo {author} {\bibfnamefont {A.}~\bibnamefont
  {Fall}}, \bibinfo {author} {\bibfnamefont {J.~D.}\ \bibnamefont {Rimstidt}},\
  and\ \bibinfo {author} {\bibfnamefont {R.~J.}\ \bibnamefont {Bodnar}},\
  }\bibfield  {title} {\enquote {\bibinfo {title} {The effect of fluid
  inclusion size on determination of homogenization temperature and density of
  liquid-rich aqueous inclusions},}\ }\href
  {https://doi.org/10.2138/am.2009.3186} {\bibfield  {journal} {\bibinfo
  {journal} {American Mineralogist}\ }\textbf {\bibinfo {volume} {94}},\
  \bibinfo {pages} {1569--1579} (\bibinfo {year} {2009})}\BibitemShut {NoStop}%
\bibitem [{\citenamefont {Roberts}\ and\ \citenamefont
  {Spencer}(1995)}]{roberts_paleotemperatures_1995}%
  \BibitemOpen
  \bibfield  {author} {\bibinfo {author} {\bibfnamefont {S.~M.}\ \bibnamefont
  {Roberts}}\ and\ \bibinfo {author} {\bibfnamefont {R.~J.}\ \bibnamefont
  {Spencer}},\ }\bibfield  {title} {\enquote {\bibinfo {title}
  {Paleotemperatures preserved in fluid inclusions in halite},}\ }\href
  {https://doi.org/10.1016/0016-7037(95)00253-V} {\bibfield  {journal}
  {\bibinfo  {journal} {Geochim. Cosmochim. Acta}\ }\textbf {\bibinfo {volume}
  {59}},\ \bibinfo {pages} {3929--3942} (\bibinfo {year} {1995})}\BibitemShut
  {NoStop}%
\bibitem [{\citenamefont {Lowenstein}, \citenamefont {Li},\ and\ \citenamefont
  {Brown}(1998)}]{lowenstein_paleotemperatures_1998}%
  \BibitemOpen
  \bibfield  {author} {\bibinfo {author} {\bibfnamefont {T.~K.}\ \bibnamefont
  {Lowenstein}}, \bibinfo {author} {\bibfnamefont {J.}~\bibnamefont {Li}},\
  and\ \bibinfo {author} {\bibfnamefont {C.~B.}\ \bibnamefont {Brown}},\
  }\bibfield  {title} {\enquote {\bibinfo {title} {Paleotemperatures from fluid
  inclusions in halite: method verification and a 100,000 year paleotemperature
  record, {D}eath {V}alley, {CA}},}\ }\href
  {https://doi.org/10.1016/S0009-2541(98)00061-8} {\bibfield  {journal}
  {\bibinfo  {journal} {Chem. Geol.}\ }\textbf {\bibinfo {volume} {150}},\
  \bibinfo {pages} {223--245} (\bibinfo {year} {1998})}\BibitemShut {NoStop}%
\bibitem [{\citenamefont {Guillerm}\ \emph {et~al.}(2020)\citenamefont
  {Guillerm}, \citenamefont {Gardien}, \citenamefont {Ariztegui},\ and\
  \citenamefont {Caupin}}]{guillerm_restoring_2020}%
  \BibitemOpen
  \bibfield  {author} {\bibinfo {author} {\bibfnamefont {E.}~\bibnamefont
  {Guillerm}}, \bibinfo {author} {\bibfnamefont {V.}~\bibnamefont {Gardien}},
  \bibinfo {author} {\bibfnamefont {D.}~\bibnamefont {Ariztegui}},\ and\
  \bibinfo {author} {\bibfnamefont {F.}~\bibnamefont {Caupin}},\ }\bibfield
  {title} {\enquote {\bibinfo {title} {Restoring halite fluid inclusions as an
  accurate palaeothermometer: Brillouin thermometry versus microthermometry},}\
  }\href {https://doi.org/10.1111/ggr.12312} {\bibfield  {journal} {\bibinfo
  {journal} {Geostand Geoanal Res}\ }\textbf {\bibinfo {volume} {44}},\
  \bibinfo {pages} {243--264} (\bibinfo {year} {2020})}\BibitemShut {NoStop}%
\bibitem [{\citenamefont {Kr{\"u}ger}\ \emph {et~al.}(2011)\citenamefont
  {Kr{\"u}ger}, \citenamefont {Marti}, \citenamefont {Staub}, \citenamefont
  {Fleitmann},\ and\ \citenamefont {Frenz}}]{kruger_liquid_2011}%
  \BibitemOpen
  \bibfield  {author} {\bibinfo {author} {\bibfnamefont {Y.}~\bibnamefont
  {Kr{\"u}ger}}, \bibinfo {author} {\bibfnamefont {D.}~\bibnamefont {Marti}},
  \bibinfo {author} {\bibfnamefont {R.~H.}\ \bibnamefont {Staub}}, \bibinfo
  {author} {\bibfnamefont {D.}~\bibnamefont {Fleitmann}},\ and\ \bibinfo
  {author} {\bibfnamefont {M.}~\bibnamefont {Frenz}},\ }\bibfield  {title}
  {\enquote {\bibinfo {title} {Liquid\textendash vapour homogenisation of fluid
  inclusions in stalagmites: Evaluation of a new thermometer for palaeoclimate
  research},}\ }\href {https://doi.org/10.1016/j.chemgeo.2011.07.009}
  {\bibfield  {journal} {\bibinfo  {journal} {Chem. Geol.}\ }\textbf {\bibinfo
  {volume} {289}},\ \bibinfo {pages} {39--47} (\bibinfo {year}
  {2011})}\BibitemShut {NoStop}%
\bibitem [{\citenamefont {Spadin}\ \emph {et~al.}(2015)\citenamefont {Spadin},
  \citenamefont {Marti}, \citenamefont {{Hidalgo-Staub}}, \citenamefont {Ri{\v
  c}ka}, \citenamefont {Fleitmann},\ and\ \citenamefont
  {Frenz}}]{spadin_technical_2015}%
  \BibitemOpen
  \bibfield  {author} {\bibinfo {author} {\bibfnamefont {F.}~\bibnamefont
  {Spadin}}, \bibinfo {author} {\bibfnamefont {D.}~\bibnamefont {Marti}},
  \bibinfo {author} {\bibfnamefont {R.}~\bibnamefont {{Hidalgo-Staub}}},
  \bibinfo {author} {\bibfnamefont {J.}~\bibnamefont {Ri{\v c}ka}}, \bibinfo
  {author} {\bibfnamefont {D.}~\bibnamefont {Fleitmann}},\ and\ \bibinfo
  {author} {\bibfnamefont {M.}~\bibnamefont {Frenz}},\ }\bibfield  {title}
  {\enquote {\bibinfo {title} {Technical note: How accurate can stalagmite
  formation temperatures be determined using vapour bubble radius measurements
  in fluid inclusions?}}\ }\href {https://doi.org/10.5194/cp-11-905-2015}
  {\bibfield  {journal} {\bibinfo  {journal} {Clim. Past}\ }\textbf {\bibinfo
  {volume} {11}},\ \bibinfo {pages} {905--913} (\bibinfo {year}
  {2015})}\BibitemShut {NoStop}%
\bibitem [{\citenamefont {{{The International Association for the Properties of
  Water and
  Steam}}}(2018)}]{theinternationalassociationforthepropertiesofwaterandsteam_revised_2018}%
  \BibitemOpen
  \bibfield  {author} {\bibinfo {author} {\bibnamefont {{{The International
  Association for the Properties of Water and Steam}}}},\ }\href
  {http://www.iapws.org/relguide/IAPWS95-2018.pdf} {\enquote {\bibinfo {title}
  {Revised release on the {IAPWS} formulation 1995 for the thermodynamic
  properties of ordinary water substance for general and scientific use},}\
  }\bibinfo {type} {Tech. Rep.}\ \bibinfo {number} {IAPWS R6-95(2018)}\
  (\bibinfo  {institution} {The International Association for the Properties of
  Water and Steam},\ \bibinfo {year} {2018})\BibitemShut {NoStop}%
\bibitem [{\citenamefont {Wagner}\ and\ \citenamefont
  {Pruss}(2002)}]{wagner_iapws_2002}%
  \BibitemOpen
  \bibfield  {author} {\bibinfo {author} {\bibfnamefont {W.}~\bibnamefont
  {Wagner}}\ and\ \bibinfo {author} {\bibfnamefont {A.}~\bibnamefont {Pruss}},\
  }\bibfield  {title} {\enquote {\bibinfo {title} {The {{IAPWS}} formulation
  1995 for the thermodynamic properties of ordinary water substance for general
  and scientific use},}\ }\href {https://doi.org/10.1063/1.1461829} {\bibfield
  {journal} {\bibinfo  {journal} {J. Phys. Chem. Ref. Data}\ }\textbf {\bibinfo
  {volume} {31}},\ \bibinfo {pages} {387--535} (\bibinfo {year}
  {2002})}\BibitemShut {NoStop}%
\bibitem [{\citenamefont {Caupin}(2005)}]{caupin_liquid-vapor_2005}%
  \BibitemOpen
  \bibfield  {author} {\bibinfo {author} {\bibfnamefont {F.}~\bibnamefont
  {Caupin}},\ }\bibfield  {title} {\enquote {\bibinfo {title} {Liquid-vapor
  interface, cavitation, and the phase diagram of water},}\ }\href
  {https://doi.org/10.1103/PhysRevE.71.051605} {\bibfield  {journal} {\bibinfo
  {journal} {Phys. Rev. E}\ }\textbf {\bibinfo {volume} {71}},\ \bibinfo
  {pages} {051605} (\bibinfo {year} {2005})}\BibitemShut {NoStop}%
\bibitem [{\citenamefont {Bruot}\ and\ \citenamefont
  {Caupin}(2016)}]{bruot_curvature_2016}%
  \BibitemOpen
  \bibfield  {author} {\bibinfo {author} {\bibfnamefont {N.}~\bibnamefont
  {Bruot}}\ and\ \bibinfo {author} {\bibfnamefont {F.}~\bibnamefont {Caupin}},\
  }\bibfield  {title} {\enquote {\bibinfo {title} {Curvature {{Dependence}} of
  the {{Liquid-Vapor Surface Tension}} beyond the {{Tolman Approximation}}},}\
  }\href {https://doi.org/10.1103/PhysRevLett.116.056102} {\bibfield  {journal}
  {\bibinfo  {journal} {Phys. Rev. Lett.}\ }\textbf {\bibinfo {volume} {116}},\
  \bibinfo {pages} {056102} (\bibinfo {year} {2016})}\BibitemShut {NoStop}%
\bibitem [{\citenamefont {{{The International Association for the Properties of
  Water and
  Steam}}}(2014)}]{theinternationalassociationforthepropertiesofwaterandsteam_revised_2014}%
  \BibitemOpen
  \bibfield  {author} {\bibinfo {author} {\bibnamefont {{{The International
  Association for the Properties of Water and Steam}}}},\ }\href
  {http://iapws.org/relguide/Surf-H2O-2014.pdf} {\enquote {\bibinfo {title}
  {Revised release on surface tension of ordinary water substance},}\ }\bibinfo
  {type} {Tech. Rep.}\ \bibinfo {number} {R1-76(2014)}\ (\bibinfo
  {institution} {{The International Association for the Properties of Water and
  Steam}},\ \bibinfo {address} {Moscow},\ \bibinfo {year} {2014})\BibitemShut
  {NoStop}%
\bibitem [{\citenamefont {Rogers}\ and\ \citenamefont
  {Pitzer}(1982)}]{rogers_volumetric_1982}%
  \BibitemOpen
  \bibfield  {author} {\bibinfo {author} {\bibfnamefont {P.~S.~Z.}\
  \bibnamefont {Rogers}}\ and\ \bibinfo {author} {\bibfnamefont {K.~S.}\
  \bibnamefont {Pitzer}},\ }\bibfield  {title} {\enquote {\bibinfo {title}
  {Volumetric properties of aqueous sodium chloride solutions},}\ }\href
  {https://doi.org/10.1063/1.555660} {\bibfield  {journal} {\bibinfo  {journal}
  {J. Phys. Chem. Ref. Data}\ }\textbf {\bibinfo {volume} {11}},\ \bibinfo
  {pages} {15--81} (\bibinfo {year} {1982})}\BibitemShut {NoStop}%
\bibitem [{\citenamefont {Dutcher}, \citenamefont {Wexler},\ and\ \citenamefont
  {Clegg}(2010)}]{dutcher_surface_2010}%
  \BibitemOpen
  \bibfield  {author} {\bibinfo {author} {\bibfnamefont {C.~S.}\ \bibnamefont
  {Dutcher}}, \bibinfo {author} {\bibfnamefont {A.~S.}\ \bibnamefont
  {Wexler}},\ and\ \bibinfo {author} {\bibfnamefont {S.~L.}\ \bibnamefont
  {Clegg}},\ }\bibfield  {title} {\enquote {\bibinfo {title} {Surface tensions
  of inorganic multicomponent aqueous electrolyte solutions and melts},}\
  }\href {https://doi.org/10.1021/jp105191z} {\bibfield  {journal} {\bibinfo
  {journal} {J. Phys. Chem. A}\ }\textbf {\bibinfo {volume} {114}},\ \bibinfo
  {pages} {12216--12230} (\bibinfo {year} {2010})}\BibitemShut {NoStop}%
\bibitem [{\citenamefont {Nayar}\ \emph {et~al.}(2014)\citenamefont {Nayar},
  \citenamefont {Panchanathan}, \citenamefont {McKinley},\ and\ \citenamefont
  {Lienhard}}]{nayar_surface_2014}%
  \BibitemOpen
  \bibfield  {author} {\bibinfo {author} {\bibfnamefont {K.~G.}\ \bibnamefont
  {Nayar}}, \bibinfo {author} {\bibfnamefont {D.}~\bibnamefont {Panchanathan}},
  \bibinfo {author} {\bibfnamefont {G.~H.}\ \bibnamefont {McKinley}},\ and\
  \bibinfo {author} {\bibfnamefont {J.~H.}\ \bibnamefont {Lienhard}},\
  }\bibfield  {title} {\enquote {\bibinfo {title} {Surface tension of
  seawater},}\ }\href {https://doi.org/10.1063/1.4899037} {\bibfield  {journal}
  {\bibinfo  {journal} {Journal of Physical and Chemical Reference Data}\
  }\textbf {\bibinfo {volume} {43}},\ \bibinfo {pages} {043103} (\bibinfo
  {year} {2014})}\BibitemShut {NoStop}%
\bibitem [{\citenamefont {Farelo}, \citenamefont {Von~Brachel},\ and\
  \citenamefont {Offermann}(1993)}]{farelo_solidliquid_1993}%
  \BibitemOpen
  \bibfield  {author} {\bibinfo {author} {\bibfnamefont {F.}~\bibnamefont
  {Farelo}}, \bibinfo {author} {\bibfnamefont {G.}~\bibnamefont
  {Von~Brachel}},\ and\ \bibinfo {author} {\bibfnamefont {H.}~\bibnamefont
  {Offermann}},\ }\bibfield  {title} {\enquote {\bibinfo {title} {Solid-liquid
  equilibria in the ternary system {{NaCl-KCl-H}}{\textsubscript{2}}{{O}}},}\
  }\href {https://doi.org/10.1002/cjce.5450710119} {\bibfield  {journal}
  {\bibinfo  {journal} {Can. J. Chem. Eng.}\ }\textbf {\bibinfo {volume}
  {71}},\ \bibinfo {pages} {141--146} (\bibinfo {year} {1993})}\BibitemShut
  {NoStop}%
\bibitem [{\citenamefont {Reiss}, \citenamefont {Mirabel},\ and\ \citenamefont
  {Whetten}(1988)}]{reiss_capillarity_1988}%
  \BibitemOpen
  \bibfield  {author} {\bibinfo {author} {\bibfnamefont {H.}~\bibnamefont
  {Reiss}}, \bibinfo {author} {\bibfnamefont {P.}~\bibnamefont {Mirabel}},\
  and\ \bibinfo {author} {\bibfnamefont {R.~L.}\ \bibnamefont {Whetten}},\
  }\bibfield  {title} {\enquote {\bibinfo {title} {Capillarity theory for the
  ``coexistence'' of liquid and solid clusters},}\ }\href {https://doi.org/doi:
  10.1021/j100337a016} {\bibfield  {journal} {\bibinfo  {journal} {J. Phys.
  Chem.}\ }\textbf {\bibinfo {volume} {92}},\ \bibinfo {pages} {7241--7246}
  (\bibinfo {year} {1988})}\BibitemShut {NoStop}%
\bibitem [{\citenamefont {{{The International Association for the Properties of
  Water and
  Steam}}}(1992)}]{theinternationalassociationforthepropertiesofwaterandsteam_revised_1992}%
  \BibitemOpen
  \bibfield  {author} {\bibinfo {author} {\bibnamefont {{{The International
  Association for the Properties of Water and Steam}}}},\ }\href
  {http://iapws.org/relguide/Supp-sat.pdf} {\enquote {\bibinfo {title} {Revised
  supplementary release on saturation properties of ordinary water
  substance},}\ }\bibinfo {type} {Tech. Rep.}\ \bibinfo {number} {IAPWS
  SR1-86(1992)}\ (\bibinfo  {institution} {The International Association for
  the Properties of Water and Steam},\ \bibinfo {year} {1992})\BibitemShut
  {NoStop}%
\bibitem [{\citenamefont {Al~Ghafri}, \citenamefont {Maitland},\ and\
  \citenamefont {Trusler}(2012)}]{alghafri_densities_2012}%
  \BibitemOpen
  \bibfield  {author} {\bibinfo {author} {\bibfnamefont {S.}~\bibnamefont
  {Al~Ghafri}}, \bibinfo {author} {\bibfnamefont {G.~C.}\ \bibnamefont
  {Maitland}},\ and\ \bibinfo {author} {\bibfnamefont {J.~P.~M.}\ \bibnamefont
  {Trusler}},\ }\bibfield  {title} {\enquote {\bibinfo {title} {Densities of
  aqueous {MgCl$_2$(aq), CaCl$_2$(aq), KI(aq), NaCl(aq), KCl(aq), AlCl$_3$(aq),
  and (0.964 NaCl + 0.136 KCl)(aq) at temperatures between 283 and 472 K,
  pressures up to 68.5 MPa, and molalities up to 6 mol kg$^{-1}$}},}\ }\href
  {https://doi.org/10.1021/je2013704} {\bibfield  {journal} {\bibinfo
  {journal} {J. Chem. Eng. Data}\ }\textbf {\bibinfo {volume} {57}},\ \bibinfo
  {pages} {1288--1304} (\bibinfo {year} {2012})}\BibitemShut {NoStop}%
\bibitem [{\citenamefont {Giacomello}\ \emph {et~al.}(2013)\citenamefont
  {Giacomello}, \citenamefont {Chinappi}, \citenamefont {Meloni},\ and\
  \citenamefont {Casciola}}]{giacomello_geometry_2013}%
  \BibitemOpen
  \bibfield  {author} {\bibinfo {author} {\bibfnamefont {A.}~\bibnamefont
  {Giacomello}}, \bibinfo {author} {\bibfnamefont {M.}~\bibnamefont
  {Chinappi}}, \bibinfo {author} {\bibfnamefont {S.}~\bibnamefont {Meloni}},\
  and\ \bibinfo {author} {\bibfnamefont {C.~M.}\ \bibnamefont {Casciola}},\
  }\bibfield  {title} {\enquote {\bibinfo {title} {Geometry as a {{Catalyst}}:
  {{How Vapor Cavities Nucleate}} from {{Defects}}},}\ }\href
  {https://doi.org/10.1021/la403733a} {\bibfield  {journal} {\bibinfo
  {journal} {Langmuir}\ }\textbf {\bibinfo {volume} {29}},\ \bibinfo {pages}
  {14873--14884} (\bibinfo {year} {2013})}\BibitemShut {NoStop}%
\bibitem [{\citenamefont {Archer}\ and\ \citenamefont
  {Carter}(2000)}]{archer_thermodynamic_2000}%
  \BibitemOpen
  \bibfield  {author} {\bibinfo {author} {\bibfnamefont {D.~G.}\ \bibnamefont
  {Archer}}\ and\ \bibinfo {author} {\bibfnamefont {R.~W.}\ \bibnamefont
  {Carter}},\ }\bibfield  {title} {\enquote {\bibinfo {title} {Thermodynamic
  properties of the {{NaCl}} + {{H}}{\textsubscript{2}}{{O}} system. 4.
  {{Heat}} capacities of {{H}}{\textsubscript{2}}{{O}} and {{NaCl}}(aq) in
  cold-stable and supercooled states},}\ }\href
  {https://doi.org/10.1021/jp0003914} {\bibfield  {journal} {\bibinfo
  {journal} {J. Phys. Chem. B}\ }\textbf {\bibinfo {volume} {104}},\ \bibinfo
  {pages} {8563--8584} (\bibinfo {year} {2000})}\BibitemShut {NoStop}%
\bibitem [{\citenamefont {{Mekki-Azouzi}}\ \emph {et~al.}(2015)\citenamefont
  {{Mekki-Azouzi}}, \citenamefont {Tripathi}, \citenamefont {Pallares},
  \citenamefont {Gardien},\ and\ \citenamefont
  {Caupin}}]{mekki-azouzi_brillouin_2015}%
  \BibitemOpen
  \bibfield  {author} {\bibinfo {author} {\bibfnamefont {M.~E.}\ \bibnamefont
  {{Mekki-Azouzi}}}, \bibinfo {author} {\bibfnamefont {C.~S.~P.}\ \bibnamefont
  {Tripathi}}, \bibinfo {author} {\bibfnamefont {G.}~\bibnamefont {Pallares}},
  \bibinfo {author} {\bibfnamefont {V.}~\bibnamefont {Gardien}},\ and\ \bibinfo
  {author} {\bibfnamefont {F.}~\bibnamefont {Caupin}},\ }\bibfield  {title}
  {\enquote {\bibinfo {title} {Brillouin spectroscopy of fluid inclusions
  proposed as a paleothermometer for subsurface rocks},}\ }\href
  {https://doi.org/10.1038/srep13168} {\bibfield  {journal} {\bibinfo
  {journal} {Sci. Rep.}\ }\textbf {\bibinfo {volume} {5}} (\bibinfo {year}
  {2015}),\ 10.1038/srep13168}\BibitemShut {NoStop}%
\bibitem [{\citenamefont {Brall}\ \emph {et~al.}(2022)\citenamefont {Brall},
  \citenamefont {Gardien}, \citenamefont {Ariztegui}, \citenamefont {Sorrel},
  \citenamefont {Guillerm},\ and\ \citenamefont
  {Caupin}}]{brall_reconstructing_2022}%
  \BibitemOpen
  \bibfield  {author} {\bibinfo {author} {\bibfnamefont {N.~S.}\ \bibnamefont
  {Brall}}, \bibinfo {author} {\bibfnamefont {V.}~\bibnamefont {Gardien}},
  \bibinfo {author} {\bibfnamefont {D.}~\bibnamefont {Ariztegui}}, \bibinfo
  {author} {\bibfnamefont {P.}~\bibnamefont {Sorrel}}, \bibinfo {author}
  {\bibfnamefont {E.}~\bibnamefont {Guillerm}},\ and\ \bibinfo {author}
  {\bibfnamefont {F.}~\bibnamefont {Caupin}},\ }\bibfield  {title} {\enquote
  {\bibinfo {title} {Reconstructing lake bottom water temperatures and their
  seasonal variability in the dead sea basin during {{MIS5e}}},}\ }\href
  {https://doi.org/10.1002/dep2.185} {\bibfield  {journal} {\bibinfo  {journal}
  {Depositional Rec.}\ }\textbf {\bibinfo {volume} {n/a}} (\bibinfo {year}
  {2022}),\ 10.1002/dep2.185}\BibitemShut {NoStop}%
\bibitem [{\citenamefont {Binder}\ \emph {et~al.}(2012)\citenamefont {Binder},
  \citenamefont {Block}, \citenamefont {Virnau},\ and\ \citenamefont
  {Tr{\"o}ster}}]{binder_van_2012}%
  \BibitemOpen
  \bibfield  {author} {\bibinfo {author} {\bibfnamefont {K.}~\bibnamefont
  {Binder}}, \bibinfo {author} {\bibfnamefont {B.~J.}\ \bibnamefont {Block}},
  \bibinfo {author} {\bibfnamefont {P.}~\bibnamefont {Virnau}},\ and\ \bibinfo
  {author} {\bibfnamefont {A.}~\bibnamefont {Tr{\"o}ster}},\ }\bibfield
  {title} {\enquote {\bibinfo {title} {Beyond the van der waals loop: What can
  be learned from simulating lennard-jones fluids inside the region of phase
  coexistence},}\ }\href {https://doi.org/10.1119/1.4754020} {\bibfield
  {journal} {\bibinfo  {journal} {American Journal of Physics}\ }\textbf
  {\bibinfo {volume} {80}},\ \bibinfo {pages} {1099--1109} (\bibinfo {year}
  {2012})}\BibitemShut {NoStop}%
\bibitem [{\citenamefont {Cohan}(1938)}]{cohan_sorption_1938}%
  \BibitemOpen
  \bibfield  {author} {\bibinfo {author} {\bibfnamefont {L.~H.}\ \bibnamefont
  {Cohan}},\ }\bibfield  {title} {\enquote {\bibinfo {title} {Sorption
  hysteresis and the vapor pressure of concave surfaces},}\ }\href
  {https://doi.org/doi: 10.1021/ja01269a058} {\bibfield  {journal} {\bibinfo
  {journal} {J. Am. Chem. Soc.}\ }\textbf {\bibinfo {volume} {60}},\ \bibinfo
  {pages} {433--435} (\bibinfo {year} {1938})}\BibitemShut {NoStop}%
\bibitem [{\citenamefont {Cole}\ and\ \citenamefont
  {Saam}(1974)}]{cole_excitation_1974}%
  \BibitemOpen
  \bibfield  {author} {\bibinfo {author} {\bibfnamefont {M.~W.}\ \bibnamefont
  {Cole}}\ and\ \bibinfo {author} {\bibfnamefont {W.~F.}\ \bibnamefont
  {Saam}},\ }\bibfield  {title} {\enquote {\bibinfo {title} {Excitation
  spectrum and thermodynamic properties of liquid films in cylindrical
  pores},}\ }\href {https://doi.org/10.1103/PhysRevLett.32.985} {\bibfield
  {journal} {\bibinfo  {journal} {Phys. Rev. Lett.}\ }\textbf {\bibinfo
  {volume} {32}},\ \bibinfo {pages} {985--988} (\bibinfo {year}
  {1974})}\BibitemShut {NoStop}%
\bibitem [{\citenamefont {Saam}\ and\ \citenamefont
  {Cole}(1975)}]{saam_excitations_1975}%
  \BibitemOpen
  \bibfield  {author} {\bibinfo {author} {\bibfnamefont {W.~F.}\ \bibnamefont
  {Saam}}\ and\ \bibinfo {author} {\bibfnamefont {M.~W.}\ \bibnamefont
  {Cole}},\ }\bibfield  {title} {\enquote {\bibinfo {title} {Excitations and
  thermodynamics for liquid-helium films},}\ }\href
  {https://doi.org/10.1103/PhysRevB.11.1086} {\bibfield  {journal} {\bibinfo
  {journal} {Phys. Rev. B}\ }\textbf {\bibinfo {volume} {11}},\ \bibinfo
  {pages} {1086--1105} (\bibinfo {year} {1975})}\BibitemShut {NoStop}%
\bibitem [{\citenamefont {Caupin}\ \emph {et~al.}(2008)\citenamefont {Caupin},
  \citenamefont {Herbert}, \citenamefont {Balibar},\ and\ \citenamefont
  {Cole}}]{caupin_comment_2008}%
  \BibitemOpen
  \bibfield  {author} {\bibinfo {author} {\bibfnamefont {F.}~\bibnamefont
  {Caupin}}, \bibinfo {author} {\bibfnamefont {E.}~\bibnamefont {Herbert}},
  \bibinfo {author} {\bibfnamefont {S.}~\bibnamefont {Balibar}},\ and\ \bibinfo
  {author} {\bibfnamefont {M.}~\bibnamefont {Cole}},\ }\bibfield  {title}
  {\enquote {\bibinfo {title} {Comment on ``{{Nanoscale}} water capillary
  bridges under deeply negative pressure'' [{{{\emph{Chem}}}}{\emph{.
  }}{{{\emph{Phys}}}}{\emph{. }}{{{\emph{Lett}}}}{\emph{.}} 451 (2008) 88]},}\
  }\href {https://doi.org/10.1016/j.cplett.2008.08.047} {\bibfield  {journal}
  {\bibinfo  {journal} {Chem. Phys. Lett.}\ }\textbf {\bibinfo {volume}
  {463}},\ \bibinfo {pages} {283--285} (\bibinfo {year} {2008})}\BibitemShut
  {NoStop}%
\bibitem [{\citenamefont {Pellegrin}\ \emph {et~al.}(2020)\citenamefont
  {Pellegrin}, \citenamefont {Bouret}, \citenamefont {Celestini},\ and\
  \citenamefont {Noblin}}]{pellegrin_cavitation_2020}%
  \BibitemOpen
  \bibfield  {author} {\bibinfo {author} {\bibfnamefont {M.}~\bibnamefont
  {Pellegrin}}, \bibinfo {author} {\bibfnamefont {Y.}~\bibnamefont {Bouret}},
  \bibinfo {author} {\bibfnamefont {F.}~\bibnamefont {Celestini}},\ and\
  \bibinfo {author} {\bibfnamefont {X.}~\bibnamefont {Noblin}},\ }\bibfield
  {title} {\enquote {\bibinfo {title} {Cavitation mean expectation time in a
  stretched {{Lennard-Jones}} fluid under confinement},}\ }\href
  {https://doi.org/10.1021/acs.langmuir.0c01886} {\bibfield  {journal}
  {\bibinfo  {journal} {Langmuir}\ }\textbf {\bibinfo {volume} {36}},\ \bibinfo
  {pages} {14181--14188} (\bibinfo {year} {2020})}\BibitemShut {NoStop}%
\bibitem [{\citenamefont {Amabili}\ \emph {et~al.}(2016)\citenamefont
  {Amabili}, \citenamefont {Lisi}, \citenamefont {Giacomello},\ and\
  \citenamefont {Casciola}}]{amabili_wetting_2016}%
  \BibitemOpen
  \bibfield  {author} {\bibinfo {author} {\bibfnamefont {M.}~\bibnamefont
  {Amabili}}, \bibinfo {author} {\bibfnamefont {E.}~\bibnamefont {Lisi}},
  \bibinfo {author} {\bibfnamefont {A.}~\bibnamefont {Giacomello}},\ and\
  \bibinfo {author} {\bibfnamefont {C.~M.}\ \bibnamefont {Casciola}},\
  }\bibfield  {title} {\enquote {\bibinfo {title} {Wetting and cavitation
  pathways on nanodecorated surfaces},}\ }\href
  {https://doi.org/10.1039/C5SM02794B} {\bibfield  {journal} {\bibinfo
  {journal} {Soft Matter}\ }\textbf {\bibinfo {volume} {12}},\ \bibinfo {pages}
  {3046--3055} (\bibinfo {year} {2016})}\BibitemShut {NoStop}%
\end{thebibliography}%
\end{document}